\documentclass[preprint]{aastex}

\slugcomment{Accepted to ApJ}

\shorttitle{Warm H$_2$O and OH disk emission in V1331~Cyg}
\shortauthors{Doppmann et al.}

\newcommand{\Msun}{M_{\odot}}

\newcommand{\Rin}{R_{\rm in}}
\newcommand{\Rout}{R_{\rm out}}

\newcommand{\Rsun}{R_{\odot}}
\newcommand{\Msunperyr}{M_{\odot}\,{\rm yr}^{-1}}
\newcommand{\Lsun}{L_{\odot}}

\newcommand{\persqcm}{\rm \,cm^{-2}}

\newcommand{\vsini}{v\sin~i}
\newcommand{\kms}{{\rm km}\,{\rm s}^{-1}}

\begin{document}


\title{Warm H$_2$O and OH disk emission in V1331~Cyg
\footnote{Data presented herein were obtained at the W.M. 
Keck Observatory from telescope time allocated to the National 
Aeronautics and Space Administration through the agency's scientific 
partnership with the California Institute of Technology and the 
University of California.  The Observatory was made possible by the 
generous financial support of the W.M. Keck Foundation.}}

%\end{center}
%\vspace{0.2cm}

\author{Greg W. Doppmann\altaffilmark{2},
Joan R. Najita\altaffilmark{2}, John S. Carr\altaffilmark{3},
 and James R. Graham\altaffilmark{4,5}}

\email{gdoppmann@noao.edu}
\email{najita@noao.edu}
\email{carr@nrl.navy.mil}
\email{jrg@berkeley.edu}


\altaffiltext{2}{National Optical Astronomy Observatory, 950 N. Cherry Ave., Tucson, AZ 85719, USA}

\altaffiltext{3}{Naval Research Laboratory, Code 7213, Washington, DC 20375, USA}

\altaffiltext{4}{Astronomy Department, UC Berkeley, Berkeley, CA 94720, USA}

\altaffiltext{5}{Dunlap Institute for Astronomy \& Astrophysics, University of Toronto, 50 St. George St., Toronto, ON, Canada M5S 3H4}


%%%%%%%%%%%%%%%%%%%%%%%%%%%%%%%%%%%%%%%%%%%%%
% ABSTRACT
%%%%%%%%%%%%%%%%%%%%%%%%%%%%%%%%%%%%%%%%%%%%%

\begin{abstract} 

We present high resolution ($R=24,000$) $L$-band spectra 
of the young intermediate mass star V1331~Cyg obtained with NIRSPEC 
on the Keck II telescope.  The spectra show strong, rich emission from 
water and OH that likely arises from the warm surface region of the circumstellar disk.
We explore the use of the new BT2 \citep{barber2006} water line list 
in fitting the spectra, and we find that it does a much better job than 
the well-known HITRAN \citep{rothman1998} water line list in the observed wavelength range and 
for the warm temperatures probed by our data. 
By comparing the observed spectra with synthetic disk emission
models, we find that the water and OH emission lines have similar
widths (FWHM $\simeq 18\,\kms$).  If the line widths are set by disk
rotation, the OH and water emission lines probe a similar range of
disk radii in this source.
The water and OH emission are consistent with thermal emission for 
both components at a temperature $\sim 1500$\,K. 
The column densities of the emitting water and OH are large, 
$\sim 10^{21}\,\persqcm$ and 
$\sim 10^{20}\,\persqcm$, respectively.
Such a high column density of water is more than adequate to 
shield the disk midplane from external UV irradiation in the 
event of complete dust settling out of the disk atmosphere, 
enabling chemical synthesis to continue in the 
midplane despite a harsh external UV environment. 
The large OH-to-water ratio is similar to expectations for 
UV irradiated disks \citep[e.g.,][]{bethell2009}, although the 
large OH column density is less easily accounted for.  

\end{abstract}

%\vspace{-5.2truein}



%%%%%%%%%%%%%%%%%%%%%%%%%%%%%%%%%%%%%%%%%%%%%
% INTRODUCTION
%%%%%%%%%%%%%%%%%%%%%%%%%%%%%%%%%%%%%%%%%%%%%

\section{Introduction}

Near-infrared ro-vibrational emission lines of water and OH 
are probes of the physical and chemical conditions in the 
warm inner regions of gaseous disks surrounding 
young stars \citep[e.g.,][]{carr2004, najita2007}.  
Water emission has been detected previously in the $K$-band  
from high accretion rate systems ($> 10^{-7}\Msunperyr$), 
ranging from low mass objects such as 
SVS~13 \citep{carr2004} and  
DG Tau \citep{najita2000}, 
to higher mass objects 
such as V1331~Cyg \citep{najita2009} 
and [BKH2005]~08576nr292 \citep{thi2005}. 
The water emission in these systems is accompanied by 
$2.3~\mu$m CO 
overtone emission that also arises from the inner disk. 
In the lower mass objects in which water emission is 
detected, the emission arises from within $\sim 0.5$\,AU. 
In all systems the emission column densities are large 
($\sim$10$^{18}$--10$^{21} \persqcm$)
and the temperature of the emitting gas is high 
($\sim$1500\,K).  

These observations complement studies of water emission made 
at mid-infrared
wavelengths with the {\it Spitzer Space Telescope} and ground-based
facilities \citep{carr2011,carr2008,pontoppidan2010a,salyk2011,salyk2008,pontoppidan2010b,knez2007}
and longer wavelength studies with the {\it Herschel Space
Telescope} \citep[e.g.,][]{sturm2010}, which measure cooler water,
probably arising at larger disk radii on average.
The mid-infrared water emission arises from systems spanning a wide
range of accretion rates, including typical T Tauri stars such as
AA~Tau \citep[][$\lesssim 10^{-8} \,\Msunperyr$]{carr2008} 
and high accretion rate T Tauri stars 
such as DR~Tau and AS~205A 
\citep[][$\gtrsim 10^{-7}\,\Msunperyr$]{salyk2008}.
Perhaps because the mid-infrared water emission can arise under a
wide range of conditions, it appears commonly in {\it Spitzer}
spectra of T Tauri stars \citep{carr2011, salyk2011,pontoppidan2010a}.

Ro-vibrational OH emission is also believed to arise from 
the inner regions of gaseous disks.  OH emission in the 
$L$-band has been previously reported from V1331~Cyg and SVS~13 
\citep{najita2007}, and from DR~Tau and AS~205A \citep{salyk2008}.
All of these sources also show H$_2$O emission in their $L$-band spectra.
OH emission has also been detected from Herbig stars; the $L$-band spectra of these 
sources is notably lacking in water emission \citep{mandell2008,fedele2011}.
The $L$-band OH emission in the Mandell sources has temperatures of 650--1000\,K, 
and column densities of  $\sim$~10$^{14}$--10$^{17} \persqcm$.

Here we present high-resolution $L$-band spectra of  
V1331~Cyg (LkHa~120), a young star in the L988 dark 
cloud complex.  Previous studies of this source have suggested 
a spectral type \citep[A8--G5,][]{kuhi1964, chavarria1981, hamann1992} 
that is intermediate between  
those of Herbig Ae stars and typical T Tauri stars. 
There are various estimates of the distance to the source 
\citep[for a summary see][]{herbig2006}, with 
extinction studies of the region toward the L988 dark 
clouds suggesting a distance of $\sim 550$\,pc.  
The accretion rate of V1331~Cyg is high.  The Br$\gamma$ 
line flux measured by \citet{eisner2007} corresponds to an 
accretion luminosity of $8.5\Lsun$ at a distance of 550\,pc, 
putting the accretion rate at $> 3\times 10^{-7}\Msunperyr$. 

As noted above, OH emission from V1331~Cyg has been 
previously reported, based on a small portion of the 
$L$-band spectrum \citep{najita2007}.  
The data presented here cover a much larger wavelength 
range, allowing us to constrain the properties of 
the molecular emission in this spectral region. 
This study complements our earlier study of CO overtone 
and water emission in the $K$-band spectrum of V1331~Cyg 
\citep[see also][]{najita2009, carr1989}.

In our previous studies of water emission, we used a water line list that was constructed from
the \citet{partridge1997} and HITRAN line lists and refined empirically in the small
portion of the $K$-band that we studied \citep{carr2004}.  Here we explore the use of the new BT2
water line list \citep{barber2006} in modeling a much larger wavelength region.

%%%%%%%%%%%%%%%%%%%%%%%%%%%%%%%%%%%%%%%%%%%%%
% Observations
%%%%%%%%%%%%%%%%%%%%%%%%%%%%%%%%%%%%%%%%%%%%%

\section{Observations} 

\subsection{Spectroscopy} 

High resolution $L$-band spectra of V1331~Cyg were taken on 11 July
2001 with NIRSPEC \citep{mclean1998}, the multi-order cryogenic
echelle facility spectrograph on Keck II. The data were obtained
by one of us (JRG) through time granted to our NASA/Keck proposal.  
The spectra cover most of the $L$-band at a resolution of  $R \equiv \lambda / \delta \lambda$ = 24,000
 ($12.5\,\kms$) using the 0.432$\arcsec$ (3-pixel wide) slit. 
The echelle orders overfill the detector in the dispersion direction, 
and therefore wavelength coverage of most of the $L$-band was achieved 
in two grating setups. 
In the first setup (hereafter
L1), the echelle and cross disperser angles were oriented at $62.2^{\circ}$
and $33.15^{\circ}$, respectively.  In the second setup (hereafter
L2), the echelle and cross disperser angles were oriented at $64.3^{\circ}$
and $33.29^{\circ}$, respectively.  Both grating setups imaged
separate portions of orders 20--25 onto the 1024 $\times$ 1024 InSb
detector array through the KL filter.  Poor telluric transmission
in orders 24 and 25 (at $3.2~\mu$m and $3.1~\mu$m) prevented our 
using any data at these grating
settings, thus our analysis used the spectra we obtained in orders
23--20 (3.27--3.88$~\mu$m, Fig. 1).

To obtain both object and sky spectra, we nodded the
telescope along the $24\arcsec$ long slit in three positions
in an ABC nod pattern, with one position centered on the
slit and the two others offset by  $-8\arcsec$ and $+8\arcsec$ 
relative to the slit center.
Single exposures of V1331~Cyg were 30 seconds $\times$ 4 co-adds
at each nod position, resulting in 12 minutes of total integration
time in $\sim 0\farcs 5$ seeing (FWHM $L$-band) within an airmass
range of 1.35--1.45.  During the observations, the instrument 
image rotator was kept in stationary mode, keeping the slit physically 
stationary and thus allowing it to slowly rotate on the sky as 
the alt-az mounted Keck II telescope
tracked.

To remove telluric absorption features, spectra of an early type star (HR 7610, A1 IV) were taken
immediately following the V1331~Cyg observations using the same two grating configurations 
successively over a similar airmass range (1.3--1.4).  Spectra of the NIRSPEC internal
continuum lamp were used for flat-fielding.  We used telluric absorption lines identified
in the spectra of our target and calibration stars for wavelength calibration.

\subsection{Data Reduction} 

Initially, bad pixels present in the raw object, telluric, and calibration frames were identified
and removed by interpolation using the {\it fixpix} algorithm within the 
REDSPEC\footnote{See http://www2.keck.hawaii.edu/inst/nirspec/redspec}
reduction package (a custom echelle reduction package developed by Prato, Kim, and McLean).
  
Standard IRAF packages \citep{massey1992, massey1997} were used to reduce the data.
Pairs of exposures taken at different nod positions closest in time were differenced then
divided by the normalized internal continuum lamp frame for adequate sky subtraction
and flatfielding.  Each echelle order on the array was then rectified using selected
bright sky emission lines (typically 9--12 lines) present in the object frame, summed over all nod positions.
Exposures taken at the same nod position were summed after accounting for any slight spatial 
offsets (to within a single pixel).

Object and telluric calibration spectra at each nod position in each echelle order were then
extracted from a spatial profile that was $\gtrsim$ 10\% of the profile peak (i.e., 3--4 pixels),
using a background aperture off the peak (e.g., $\pm$ 8--28 pixels) for residual background 
correction.  Each extracted object and telluric spectrum was wavelength calibrated using selected 
telluric absorption lines identified from the HITRAN database \citep{rothman1998}.

Telluric features in each of the three nod positions of V1331~Cyg were removed by dividing
by the spectrum of the telluric standard taken at the same nod position along the slit.
An optimal telluric cancellation in V1331~Cyg was achieved using the IRAF package {\it telluric},
which allowed the strengths of the telluric lines to be scaled to account for a slight
difference in the airmass of the hot standard star before normalization and division with the target spectra.

We put our V1331~Cyg spectra on a flux scale using photometry from the literature.   
We constructed an SED for V1331~Cyg, using the near-IR and optical photometry of 2MASS
and \citet{chavarria1981}, respectively.
The 2MASS $K$-band photometric point is in good agreement with the 
flux calibrated $K$-band spectra from \citet{najita2009}.  
We fit the near-IR photometry with a low order function to interpolate the expected flux 
in different $L$-band orders. 
We then rescaled our NIRSPEC observations so that the flux average in each
observed order matched the interpolated value from the SED 
($\sim$~0.35--0.41\,Jy across 3.291--3.854~$\mu$m).

%%%%%%%%%%%%%%%%%%%%%%%%%%%%%%%%%%%%%%%%%%%%%
% DATA Collection
%%%%%%%%%%%%%%%%%%%%%%%%%%%%%%%%%%%%%%%%%%%%%

\section{Analysis} 

The resulting $L$-band spectra of V1331~Cyg reveal emission
features that are strong, numerous, and relatively
narrow (Fig.~1).  The narrow line widths are an advantage in 
deconstructing such a rich emission spectrum. 
The coincidence between the observed emission features and the wavelengths of 
lines of OH (red ticks) and water (blue ticks) suggest that the observed spectra are 
dominated by emission from water and OH. 
As we show below, the spectra are consistent with 
OH and water emission from the atmosphere of 
the inner disk of V1331~Cyg (Figs.~2a--h).  From the line widths and
strengths, we characterize the physical conditions in the disk where
the water and OH emission arises.

\subsection{Disk Emission Model}

The radial velocity of the observed features has a topocentric value
of --27.6 $\kms$, in agreement  with the radial velocity of the C$^{18}$O (2-1) 
emission from the envelope  which provides
an estimate of the radial velocity of the system \citep[v$_{\rm
LSR}=-1.3\,\kms$,][]{mcmuldroch1993}.  
The lack of
a velocity offset with respect to the systemic velocity is consistent
with an origin for the emission in a rotating circumstellar disk.

We therefore modeled the observed spectra of V1331~Cyg by combining 
contributions from the stellar continuum, a disk veiling continuum, 
and emission lines of H$_2$O and OH from a disk atmosphere.  
The disk line emission model, which follows the approach described in
\citet{carr2004},
assumed the geometry of a plane parallel slab
of gas in a Keplerian disk that is in vertical hydrostatic equilibrium.
The parameters describing the slab are
the gas temperature ($T$), column density ($N$), and
intrinsic line broadening,
the inner and outer radii of the emission ($\Rin$, $\Rout$),
and the line-of-sight disk rotational velocity at the inner radius
of the emission ($v \sin i$).

The molecular emission spectrum can be used to constrain these parameters 
in the following schematic way.
The relative strengths of the emission features constrain the temperature
($T$) and column density ($N$) of the emission.  The large number
of transitions in the spectrum and the wide range in their intrinsic
strengths ($gf$-values) and energy levels are assets in this regard. 
The flux level of the emission spectrum
constrains the projected emitting area for the emission, which is
a function of $\Rin$ and $\Rout$, the inclination, and the assumed 
distance to the system.

The V1331~Cyg emission lines are resolved, narrow, and centrally peaked. 
Because the emission lines are resolved, we have a 
further constraint on the emitting radii. 
The maximum velocity of the emission profile 
(i.e., the velocity extent of the profile) constrains 
the projected velocity at the inner radius of the emission 
($v\sin i$).  
The values of $\Rin$ and $v\sin i$, along with an assumed stellar mass,
specify the inclination of the system.

The shape of the emission profile constrains the outer radius of
the emission ($\Rout$), e.g., a small ratio of $\Rout/\Rin$ produces a
double-peaked profile, whereas a large ratio produces a
centrally-peaked profile.
In modeling the spectrum we find that the
centrally peaked profile cannot be fit with $\Rout/\Rin < 2.0$;
therefore the emission must arise over a range of radii
and not from a narrow annulus.

The temperature and column density may also vary with radius 
and can affect the detailed line profile shape.  Because 
the line profiles are only marginally resolved in our data, our 
ability to diagnose these properties is limited.  We therefore 
assume a constant temperature and column density for 
the emitting region in the modeling carried out here. 

Thus our strategy was to search for a combination of temperature
and column density that fit the relative line strengths and then
choose an emitting area (i.e., an outer radius) that matched the
overall flux of the lines for the assumed distance.
The combination of an observed $v \sin i$, an emitting area inferred 
from the emission flux, and a constraint on $\Rout/\Rin$ 
from the line profile then allowed us to constrain the model 
parameters $\Rout$, $\Rin$, and the inclination $i$. 
We required that the model provide a decent fit to all four of
the orders observed with the L1 and L2 grating settings.

The additional parameters in the modeling are the stellar mass and 
the distance and extinction to the source.  For consistency with
our earlier study of the $K$-band water and CO emission
from this source \citep{najita2009}, we adopted
a distance of 550 pc \citep{shevchenko1991, alves1998},
a stellar mass of $1.8\Msun$,
and no $K$-band extinction along the line of sight.
Also consistent with our earlier study, we assumed a stellar radius
and temperature of $2.2\Rsun$ and $7200$\,K, respectively,
and we estimated the underlying stellar $L$-band continuum assuming a
blackbody function.  As noted in \citet{najita2009}, the stellar properties are not well known.

We constructed separate syntheses for the OH and water emission, and then
combined the models additively.  This is an acceptable approach
because there is little overlap between the lines from the water
and OH line lists to within several $\kms$, despite the high density
of water lines in the BT2 list.
To determine the strength of the disk veiling continuum, we
started with the stellar contribution to the
$L$-band continuum and added the synthesized water emission spectrum.
Disk continuum emission was then added to bring the model emission
up to the observed level of the emission in each order.
The required amount of disk continuum emission
increased monotonically from $\sim 0.26$\,Jy to $\sim 0.37$\,Jy across
$L$-band orders 23 to 20, consistent with the SED shape of V1331~Cyg
($\S$2.2).  Since most of the $L$-band continuum is dominated by the excess disk
emission (i.e., veiling of approximately 10 in the $L$-band), no stellar spectral
features would be detectable in our data.
As the last step, a synthetic OH emission spectrum was
added to produce the final model spectrum (Figs.\ 2a--h).

\subsection{Molecular Linelists}

A successful model requires good line lists.  
Our OH line list comes from the HITRAN database \citep{rothman1998}, 
an empirically determined list based on absorption in the
Earth's atmosphere.  Many of the OH emission lines in our $L$-band
spectra are prominent and mostly isolated from one another, allowing
us to identify the transitions present in our data.

For water, we use a new theoretical line list, BT2 \citep{barber2006},
which includes over 500 million transitions.  We trimmed the line list to a parameter
range that is relevant for the present application, by sorting lines based on their relative strengths at 1500K,
the approximate temperature we found in fitting the $K$-band water emission of V1331~Cyg.
Here the relative strength is the product of the absorption cross section and the fractional population in the
lower energy level of the transition.  We excluded lines that had relative strengths $< 0.001\%$ of the
strongest line in each order.  The resulting line list included $\sim$4600 to $\sim$6700 lines in each of 
our $L$-band orders.

\subsection{Water and OH emission in the $L$-band}

We find that we can fit the water emission line profile and
reproduce much of the water emission structure we observe
with a temperature of $\sim 1500$\,K,
a line-of-sight H$_2$O column density of $\sim 2 \times 10^{21} \persqcm$,
an emitting area extending from $\Rin = 5.5\Rsun$ to $\Rout = 3.5 \Rin$
(i.e., 0.03--0.09\,AU),
and $v\sin i = 14\,\kms$ at $\Rin$.
The resulting emission profile is convolved with the 3-pixel NIRSPEC
slit resolution of $12.5\,\kms$.
We refer to the model with these parameters as our ``reference model'' for the water emission (Figs.\ 2a--h).
As the fits demonstrate, the line emission is dominated by water in all observed $L$-band orders.  
The water lines detected in the various $L$-band orders span a broad range from optically thin to thick
(e.g., $\tau_{H_2O} \sim10^{-3}$ to $10^{2}$ for the reference model).  

These values for the water temperature and
column density are similar to the fit we found previously to the $K$-band
water emission observed in V1331~Cyg \citep{najita2009}.  The emitting
area we derive for the $L$-band water emission is somewhat smaller than
($\sim$~50\% of) the size of the $K$-band emitting area found in the
earlier study.
This difference does not mean that the results are in conflict, because
the $K$- and $L$-band observations were taken at different epochs and the
line emission strength may vary.
But it is also unclear whether variability is indicated because the
present data were flux calibrated using data from the literature
rather than contemporaneous observations ($\S$2).

An inner radius of 5.5$\Rsun$ is 2.5 times larger than
the stellar radius of 2.2$\Rsun$ ($\S$3.1).
For a stellar mass of $1.8\Msun$ ($\S$3.1), the adopted values of
$v\sin i$ and $\Rin$ correspond to an inclination of 3 degrees.
The inclination and the range of emission radii are
consistent with
the values we found in modeling the $K$-band water
emission from this source \citep{najita2009}.

The inferred values for $\Rin,$ $\Rout$, and inclination depend on 
the assumed distance to the source and the stellar mass, neither of 
which are well known.  The distance has been variously estimated 
in the range 500--800\,pc (Najita et al.\ 2009), allowing for larger 
values than we assumed.  A larger distance would require a larger 
disk emitting area to produce the same flux.  For those larger disk 
emitting radii, a larger inclination would be needed to produce 
the same $v \sin i$ at a fixed stellar mass. 
However, a larger distance would also affect the estimate of the 
stellar mass based on the SED, with further consequences for 
the $\Rin$, $\Rout$, and inclination.  Because of these uncertainties, 
the values of $\Rin$, $\Rout$, and inclination are not well 
constrained by our modeling.  
Nevertheless, the modeling shows that the notional values 
of stellar mass and distance from the
literature do allow for a simple disk model for the molecular emission.

To estimate the range of column density (or temperature) that would produce a reasonable fit to the
water emission, we held temperature (or column density) fixed at the value of the reference model and
allowed column density (or temperature) to vary with emitting area as a free parameter.  The emitting area 
was determined by fitting the spectrum in L1 order 21, and the value was kept the same for all other orders. 
L1 order 21 was chosen since it contained a mix of optically thick and thin lines and is a spectral region with 
relatively good telluric transmission.  We determined the value of the emitting area by minimizing  the 
sum of the absolute values of the pixel differences between the observed spectrum and the model in L1 order 21 \citep{branham1982}.  Regions with 
poor telluric transmission (i.e., $<$~80\%) were excluded in this estimate.

Figure 3a shows the resulting fits to L1 order 21 at a fixed temperature (T=1500\,K) for the reference model 
(middle spectrum) and for column densities 10 times lower (upper spectrum) and 10 times higher (lower 
spectrum).  The low column density model is more optically thin and has larger peak-to-trough variation, 
while the high column density model has much less contrast.  The peak-to-trough variation
in the reference model more closely resembles that of the observed spectrum.  Figure 3b shows the model 
results for a neighboring order (L1 order 22) obtained with the same emitting area used in L1 order 21.  The 
low and high column density models fit much worse than in L1 order 21 because the emitting area was not 
optimized for this order.  These model fits are clearly worse than the reference model in fitting the
observed spectrum.

Figure 4a shows fits to L1 order 21 for the reference model temperature (middle spectrum) and for lower 
(upper spectrum) and higher (lower spectrum) temperatures.  The lower temperature model has too large a 
peak-to-trough variation, while the higher temperature model has too little, compared with the reference 
model case (1500\,K).  This same trend is evident in another order (L1 order 22, Fig. 4b).  In both figures, 
the high and low temperature models are clearly worse than the reference model case.  Therefore, for the 
water emission it appears that we can rule out column densities $\le 2 \times 10^{20}$ and $\ge  2 \times 10^
{22} \persqcm$ and temperatures $\le 1200$\,K and $\ge 2000$\,K.

We adopted a similar approach in fitting the OH emission. 
We found that the OH emission lines in our $L$-band data can be fit 
with a temperature of $\sim 1500$\,K and a line-of-sight
OH column density of $\sim 1 \times 10^{20} \persqcm$, 
and a larger emitting area (2.1 times larger) than
the water emission, 
with radii extending from $\Rin = 5.5\Rsun$ to $\Rout = 5 \Rin$, 
and $v\sin i = 14\,\kms$ at $\Rin$.
We refer to the model with these parameters as our ``reference model'' for the OH emission (Fig.\ 2a--h).
The difference between the emitting
areas of water and OH in the reference model is not significant.
If we assume the OH emission comes from the same range of radii as the water emission  
($\Rin = 5.5\Rsun$ to $\Rout = 3.5 \Rin$), then we find a similar
fit to the OH emission at the same column density as above, 
but at a slightly higher temperature ($1850$\,K).
The OH lines in our spectra range from optically thin to thick, 
depending on the transition and model parameters (e.g., $\tau_{OH}  \sim0.1$ to $30$ for
the reference model, Table 1).  The OH transitions also span a range in excitation, 
including 1--0 lines with J = 11,~12,~14-18 and
2--1 lines with J = 8-18 (red ticks, Fig.\ 2a--h, see Table 1).   Since the OH lines in our data
span a range of optical depths and excitation levels, we can constrain the column density
and temperature of the model fits.  

By requiring a consistent fit to all orders we can exclude column densities $\le 5 \times 10^{19}$
and $\ge 10^{21} \persqcm$ and temperatures $\le 1200$\,K and $\ge 2300$\,K.
Figure 5a compares observed spectral regions where OH is present (black histogram) with model fits that
include H$_2$O and OH and adopt higher and lower OH
column densities than the reference OH model at a fixed temperature (T=1500\,K).  The top panel compares
the reference model (N=$1 \times 10^{20} \persqcm$, solid blue line) with a low column density fit (N=$5 
\times 10^{19} \persqcm$; dotted red line).  The low column density model overpredicts the flux of the lines 
at shorter wavelengths (3.282-3.287 $\mu$m).  The bottom panel compares the reference model (N=$1 
\times 10^{20} \persqcm$, solid blue line) with a high column density fit (N=$1 \times 10^{21} \persqcm$, 
dotted red line).  The high column density model underpredicts the flux of 
the lines at shorter wavelengths (3.282-3.287 $\mu$m) and somewhat overpredicts the flux
of the lines at longer wavelengths (3.776-3.780 $\mu$m).  The reference model fits all the OH lines 
adequately well across a range of optical depths.

Similarly, Figure 5b compares selected OH lines (black histogram) with fits at higher and lower
temperatures than the reference OH model at a fixed column density (N=$1 \times 10^{20} \persqcm$).
The top 
panel compares the reference model (T=1500\,K, solid blue line) with a low temperature fit (T=1200\,K; 
dotted red line).  The low temperature model overpredicts the flux of the low-J lines at shorter wavelengths 
(3.282-3.287 $\mu$m).  The bottom panel compares the reference model (T=1500\,K, solid blue line) with a 
high temperature fit (T=2300\,K, dotted red line).  The high temperature model overpredicts the flux of 
the high-J lines at longer wavelengths (3.776-3.780 $\mu$m).  The low temperature model puts too much 
energy into the low-J lines (shorter wavelengths, top panel), and the high temperature model puts too much 
energy into the high-J lines (longer wavelengths, bottom panel). The reference model fits all the OH lines
adequately well across a range of excitation levels.

The good agreement between the observed spectrum and the OH+H$_2$O model spectrum (Fig. 2)
shows that it is possible to account for much of the spectral structure
with our reference model of molecular emission from a rotating 
disk.  The temperature of the water and OH emission fit by the reference models is the same 
($\sim1500$\,K for both), while the OH column is $\sim 10$ times less than the water
column.
The range of emitting radii for both models overlaps, 
with the emission from both species arising from within 0.2\,AU 
of the star.
Some discrepancies between the observed and model spectra are noticeable and
may be attributed to errors in the transition frequency and/or oscillator 
strength of some lines in the BT2 water line list (e.g., 
features at 3.485~$\mu$m, 3.492~$\mu$m, 3.499~$\mu$m, 
3.505~$\mu$m, 3.514~$\mu$m; vertical arrows in Fig.\ 2c).

The advantage of using the BT2 list, compared to the more familiar 
HITRAN list, when studying warm water emission is apparent if we 
compare synthetic spectra produced with the two line lists 
at a temperature of 1500\,K (Fig.~6). 
While the fit parameters are identical in both cases, the BT2 list clearly matches
most of the observed structure in the data, while the HITRAN list
misses most of it.  This is because of the absence of high temperature
lines in HITRAN, which are present in the BT2 list.  
For water emission at lower temperatures ($\sim1000$\,K)
and in wavelength regions dominated by low excitation lines, 
the HITRAN list can do about as well as BT2, as in the case of the 
$2.9~\mu$m water emission observed in AS~205A and DR~Tau \citep{salyk2008}.

The line widths of individual OH and H$_2$O lines are difficult to measure directly
from our data owing to the crowding of numerous H$_2$O lines throughout the $L$-band.
In the absence of microturbulent broadening, the intrinsic line widths of the molecular emission
we detect in our observations is determined by
pure thermal broadening, which is $\sim$1 $\kms$ in our case for both H$_2$O and OH at 1500K, and
0.8 $\kms$ for CO at 1800\,K. The modeling results show that the $L$-band ro-vibrational water 
and OH emission in this source have similar emission line profiles (FWHM $\sim$~17--20 $\kms$),
indicating that emission lines in both species arise from a similar 
range of disk radii. 
Thus it seems plausible that the two species probe similar gas 
temperatures.  
If this is the case, it implies that the ratio of OH and water 
column densities is $\sim 0.1$.

\subsection{Water and CO emission in the $K$-band}

Water and CO emission from V1331~Cyg has been previously reported in the $K$-band \citep{najita2009}, 
based on an observation made in 1999, two years earlier than the observation date of the $L$-band data 
reported here. To determine how the OH and water emission properties in the $L$-band compare with those 
of CO and water reported in the $K$-band study, we re-analyzed the $K$-band data 
using the same simple slab model 
described in $\S$3.1.  Our analysis here differs in several ways from the previous analysis \citep{najita2009} 
in that we report a single fixed temperature and column density for each species rather than using radial 
gradients, and chemical equilibrium is not assumed for either species.  Our analysis also uses the new BT2 
line list, compared with a previous water line list that was empirically calibrated but for a limited region near 
the 2--0 CO bandhead \citep[i.e.,][]{carr2004}. 

We find a good fit to the water features in the $K$-band using the same parameters found for the $L$-band 
(i.e., $T_{\rm H_2O}=1500~$K and $N_{\rm H_2O}=2 \times 10^{21} \persqcm$), but with an emitting area 
that is 2 times larger (i.e., $R=5.5-26~\Rsun$).  It is possible that the different values we find for the emitting 
area of water in V1331~Cyg over different observational epochs result from real time variability in the 
emission, but some fraction of the difference could also be accounted for by uncertainties in the flux 
calibration in either epoch.  Guided by the previous fit values for CO, we found we could fit the CO emission 
reasonably well with a temperature of $T_{\rm CO}=1800$~K and a line-of-sight column density 
$N_{\rm CO}=6\times 10^{21} \persqcm$, that emits over disk radii  2.3--18 $\Rsun$, 
with $\vsini = 23\,\kms$ at $\Rin$ and 
turbulent broadening of $v_{\rm turb}=4\,\kms$ (Fig.\ 7).
The BT2 list does a good job reproducing most of the emission structure observed blueward of the CO 
bandhead (Fig.\ 7), and discrepancies in the fit to the data may illustrate the need for empirical calibration of 
its line strengths and transition energies in the $K$-band as well.  The inner radius for the emission may 
appear uncomfortably 
close to the stellar radius ($2.2\Rsun$; \S 3.1).  We note 
that the inferred disk emitting radii are not well constrained 
by our modeling and larger emitting radii are possible, in principle, 
with a larger source distance (\S 3.3).  A more complete 
modeling effort that explores the uncertainty in the source distance 
is needed to explore this possibility.

From our analysis of the $K$-band data, we find a CO-to-water column density 
ratio of $\sim$~3--10,  
consistent with the results of \citet{najita2009}. 
The larger value would apply if the water emission has the same value of 
$v_{\rm turb}=4\,\kms$ as the CO emission.
The smaller value would apply if the water emission has no turbulent 
broadening, as in the $L$-band modeling. 
The CO and water emission may experience different amounts of turbulent 
broadening if they are produced at different heights or radii in the disk atmosphere. 
We compare in Table 2 the results of the $K$-band modeling with the properties of the reference model
for the $L$-band modeling.  As shown in Table 2, the properties of the water emission observed 
in the $K$- and $L$-bands are quite similar.

 \section{Discussion}

\subsection{Comparison to other sources with disk emission}

To date, gaseous water and OH emission have been reported 
in the near-IR spectra of only a handful of young stars 
\citep{carr2004, najita2000, thi2005, salyk2008, mandell2008, najita2009, fedele2011}.  
As noted in $\S$3.3, the water emission temperature ($\sim 1500$\,K) 
and column density ($\sim 10^{21}\,\persqcm$)
that we find for V1331~Cyg are similar to the water emission 
properties reported for V1331~Cyg based on its $K$-band 
spectrum \citep{najita2009}.  
These properties are also similar to those found for the 
near-IR water emission detected from SVS~13 \citep{carr2004},
for which the radii of emission are comparably small ($\lesssim 0.3$~AU).
The water emission detected from the disk surrounding the 
high mass star 08576nr292 ($6\Msun$) was also found to have a similar 
temperature but a significantly lower column density ($\sim 10^{18}\,\persqcm$) emitting over a larger area
from an inner radius of 2~AU to an outer radius of 4~AU \citep{thi2005}.

In comparison to the T Tauri stars, AS~205 and DR~Tau \citep{salyk2008},
which also show water emission in the $L$-band, 
the emission spectrum of the more extreme T Tauri star V1331~Cyg, reported here,
is much richer (i.e., has a higher density of emission lines), probably as a consequence of 
its higher water emission temperature and column density.  
Its spectrum is similar to that of SVS~13, which was previously reported by \citet{najita2007}.

%DR~Tau and AS~205A  also show water and OH emission 
%in the $L$-band, with the gas emitting further from the star 
%($\gtrsim$~0.4~AU) 
%at a lower temperature ($\sim 1000$\,K) and 
%column density ($N_{\rm H_2O} \sim 7\times 10^{17}\persqcm$ and 
%$N_{\rm OH} \sim 2\times 10^{17}\persqcm$) than is found for 
%V1331~Cyg (Table 3).  

%% Suggest to go back to original version, which is more neutral.
%%% This looks fine to me - greg

%The OH-to-water column density ratio we find for V1331~Cyg   
%($\sim 0.05$) is lower than the ratio ($\sim$~0.3) found for 
%DR~Tau and AS~205A \citep{salyk2008}. 
%Similarly, the CO-to-water column density ratio of V1331~Cyg 
%is low ($\sim$~3) compared with the ratio of $\sim$~10 found for 
%both DR~Tau and AS~205A. {\it Add text as per John's 2nd email comment \#3}

The Herbig Ae stars AB~Aur and MWC~758 \citep{mandell2008} 
also show OH emission in the $L$-band, 
but no water emission even at a signal-to-noise of $\sim 1000$. 
\citet{mandell2008} find a temperature of $\sim 700$\,K for 
the (optically thin) OH emission that they detect. 
Assuming that the gas is collisionally excited (and in LTE) 
and that the emitting region is constant in column density 
and extends out to $\sim 15$\,AU, 
consistent with Herbig Ae disk atmosphere model of \citet{kamp2005},  
the OH mass that they infer corresponds to a low OH column density 
of $\sim 10^{15}\persqcm$.  
Lower column densities are inferred if the OH emission arises 
through fluorescence. 

At {\it Spitzer} IRS wavelengths (10--40~$\mu$m),
prominent water and OH emission is common among T Tauri stars but rare among 
Herbig stars \citep{pontoppidan2010a, carr2011}. 
Where it has been studied in detail, the 
water and OH emission detected with {\it Spitzer} 
is also characterized by larger radii and more modest temperatures 
and column densities than those found for V1331~Cyg.  
T~Tauri water and 
OH emission temperatures are $\gtrsim 500$\,K 
\citep{carr2008, carr2011, salyk2008, salyk2011}, with column densities 
of $\sim 10^{18}\,\persqcm$ for H$_2$O and 
$10^{15} - 10^{17}\,\persqcm~$\citep[][respectively]{salyk2011,carr2011} for OH. 
The OH-to-water column density ratio of $\sim 0.13$ for the typical 
T Tauri star AA Tau \citep{carr2008}
is similar to the ratio of $\sim 0.1$ found for V1331~Cyg. 
Thus, the temperature and, in particular, the column densities of 
the OH and water emission from V1331~Cyg are 
at the upper end among sources with detected water and/or OH emission. 

Interestingly, V1331~Cyg does not show strong water emission at {\it Spitzer} wavelegnths \citep{carr2011}
indicating that the disk atmosphere at larger radii than is probed by the $L$-band does not produce
much water emission.  Perhaps this is because little grain settling has taken place in the V1331~Cyg
disk atmosphere at such radii, a situation that would limit line emission from the disk atmosphere
\citep{salyk2011}.

\citet{bethell2009} have recently pointed out that 
because the UV photoabsorption cross-sections of OH and water are  
approximately continuous at $10^{-17} - 10^{-18} {\rm cm}^2$
\citep{yoshino1996, vandishoeck1984}, 
OH and water column densities of 
$\sim 10^{17}-10^{18}\persqcm$ can begin to shield the disk midplane 
from UV irradiation in the event of complete dust settling out 
of the disk atmosphere, enabling chemical synthesis to continue 
in the midplane despite a harsh external UV environment. 
The high column densities of OH and water that are observed 
in V1331~Cyg and other sources are more than adequate 
for this purpose.

\subsection{Disk Photochemistry}
 
In the analysis of spectrally unresolved spectra \citep[e.g., as observed
with {\it Spitzer} IRS,][]{carr2008}, the assumption is 
sometimes made that
molecular species with similar excitation temperatures probe the
same region of the disk.  With that assumption, one can ratio the
column density measured for each molecular species as a spatial
average over the entire emitting region and obtain a rough estimate
of the molecular abundances of the emitting gas.  
The derived abundances are uncertain, in part, because it is unclear 
whether the emission in the different molecular species arise 
from the same region of the disk.  The emission in the different 
species may arise at different radii or vertical heights in the 
disk atmosphere. 
With high resolution
spectra, as employed here, we can probe the column density of the 
molecular emission as a function of disk radius for each molecular
species, thereby providing a more robust way to probe the molecular
abundances of the emitting gas.

As described in section 3, we found that the $L$-band OH and water
emission from V1331~Cyg probe a similar range of radii.  \citet{salyk2008}
found a similar result in their $L$-band spectroscopy
of the T Tauri disks DR~Tau and AS~205A.  
Ratioing the column densities found for the OH and water emission from 
these sources yields column density ratios of OH/H$_2$O 
$\sim$~0.05--0.3.  

Thermal-chemical models of inner disk atmospheres that do not include 
UV irradiation typically predict much smaller OH/H$_2$O abundance ratios 
\citep[$\sim 10^{-3}$,][]{glassgold2004, glassgold2009}. 
Including UV irradiation of the disk, which can photodissociate 
water to produce OH, can lead to OH/H$_2$O ratios more similar 
to those observed \citep{bethell2009, glassgold2009, willacy2009}. 
Perhaps UV irradiation is also responsible for the more extreme 
ratio of OH/H$_2$O found by \citet{mandell2008} in their $L$-band 
spectroscopy of AB~Aur and MWC~758, 
where a low column density of OH is detected in emission, but no 
water emission is detected. 
UV irradiation has been invoked to account for the water emission 
properties measured for the $6\,\Msun$ source 08576nr292 
\citep{thi2005}. 
\citet{pontoppidan2010a} also discuss the possible role of UV 
photodissociation in accounting for the low equivalent width 
of OH and water emission (often consistent with non-detection)
in {\it Spitzer} spectra of Herbig AeBe stars. 

In the scenario explored by \citet{bethell2009}, 
a disk irradiated by a stronger UV field will enhance the column 
density of OH in the disk atmosphere by photodissociating water.  
The maximum OH column density is predicted to 
be limited to that where the OH optical depth to FUV photons 
is close to unity or an OH column density of $\sim2\times 10^{17}\,\persqcm$. 
Since water can heat disk atmospheres through the absorption of 
FUV photons as well as provide a major 
coolant for disk atmospheres \citep{pontoppidan2010a}, it may 
regulate the thermal structure of disk atmospheres where it is 
abundant.  \citet{bethell2009} suggest that the water 
emitting layer may therefore be limited to a column density at 
which water is optically thick to FUV photons, i.e., to a water 
column density of $<10^{18}\,\persqcm$.   
These OH and water column densities are similar to those measured for 
some T Tauri disks \citep{carr2008,salyk2008}.  

The much larger column densities of both water and OH that 
are found for V1331~Cyg and SVS~13, compared to these predictions, suggest 
that additional processes beyond FUV heating and simple photochemistry 
play a role in the thermal-chemical properties of disk atmospheres. 
Since both V1331~Cyg and SVS~13 are high accretion rate sources, 
one possibility is that non-radiative (accretion-related) mechanical heating 
deepens the disk temperature inversion region and leads to  
larger column densities of molecular emission 
compared to more typical T Tauri stars. 
In models of disk atmospheres of typical T Tauri stars,
accretion-related mechanical heating has been found to enhance 
the column density of warm water at the disk surface up to the 
level observed with {\it Spitzer} at AU distances from the star 
\citep{glassgold2009}.  The models predict 
much larger column densities of warm water $\lesssim 0.25$\,AU
than at 1 AU (Najita et al. in prep.) and may explain, in part, the large water column we observe.

Another possibility is that at the higher average temperatures 
(and smaller disk radii, higher densities) from which the 
V1331~Cyg and SVS~13 emission originates, there are additional 
ionization or chemical processes that lead to enhanced OH and 
water column densities. 
Whether these, or other processes, can account for the water 
and OH column densities observed here is an important issue for 
the future. 

\acknowledgments

The authors wish to recognize and acknowledge the very significant cultural role and reverence that the 
summit of Mauna Kea has always had within the indigenous Hawaiian community.  We are most fortunate to 
have the opportunity to conduct observations from this mountain.  We thank the Keck Observatory staff who 
provided support and assistance during our NIRSPEC run.  We thank the anonymous referee for a thorough 
reading of the manuscript, whose comments helped improve the paper.  Financial support for GWD was 
provided in part by the NASA Origins of Solar Systems program NNH10A0061.   JRG was supported in part 
by the University of California Lab Research Program 09-LR-01-118057-GRAJ and NSF AST-0909188.  JRN 
and JSC acknowledge support from the NASA Origins of Solar Systems program.
JSC also acknowledges 6.1 funding for basic research in infrared astronomy at the
Naval Research Laboratory.

{\it Facility:} \facility{Keck:II(NIRSPEC)}.

\clearpage

%%%%%%%%%%%%%%%%%%%%%%%%%%%%%%%%%%%%%%%%%%%%%
% References
%%%%%%%%%%%%%%%%%%%%%%%%%%%%%%%%%%%%%%%%%%%%%

\clearpage

%% the file, uncomment the \plotone, \plottwo, and
%% If you need a layout that cannot be achieved with \plotone or
%% \plottwo, you can invoke the graphicx package directly with the
%% \includegraphics command or use \plotfiddle. For more information,
%% please see the tutorial on "Using Electronic Art with AASTeX" in the
%% documentation section at the AASTeX Web site,
%% http://www.journals.uchicago.edu/AAS/AASTeX.
%%
%% The examples below also include sample markup for submission of
%% supplemental electronic materials. As always, be sure to check
%% the instructions to authors for the journal you are submitting to
%% for specific submissions guidelines as they vary from

%% This example uses \plotone to include an EPS file scaled to
%% 80% of its natural size with \epsscale. Its caption
%% has been written to indicate that additional figure parts will be
%% available in the electronic journal.

% FIGURES &  CAPTIONS

% Fig 1 Observed Spectra of V1331~Cyg
\begin{figure}
\figurenum{1}
\plotone{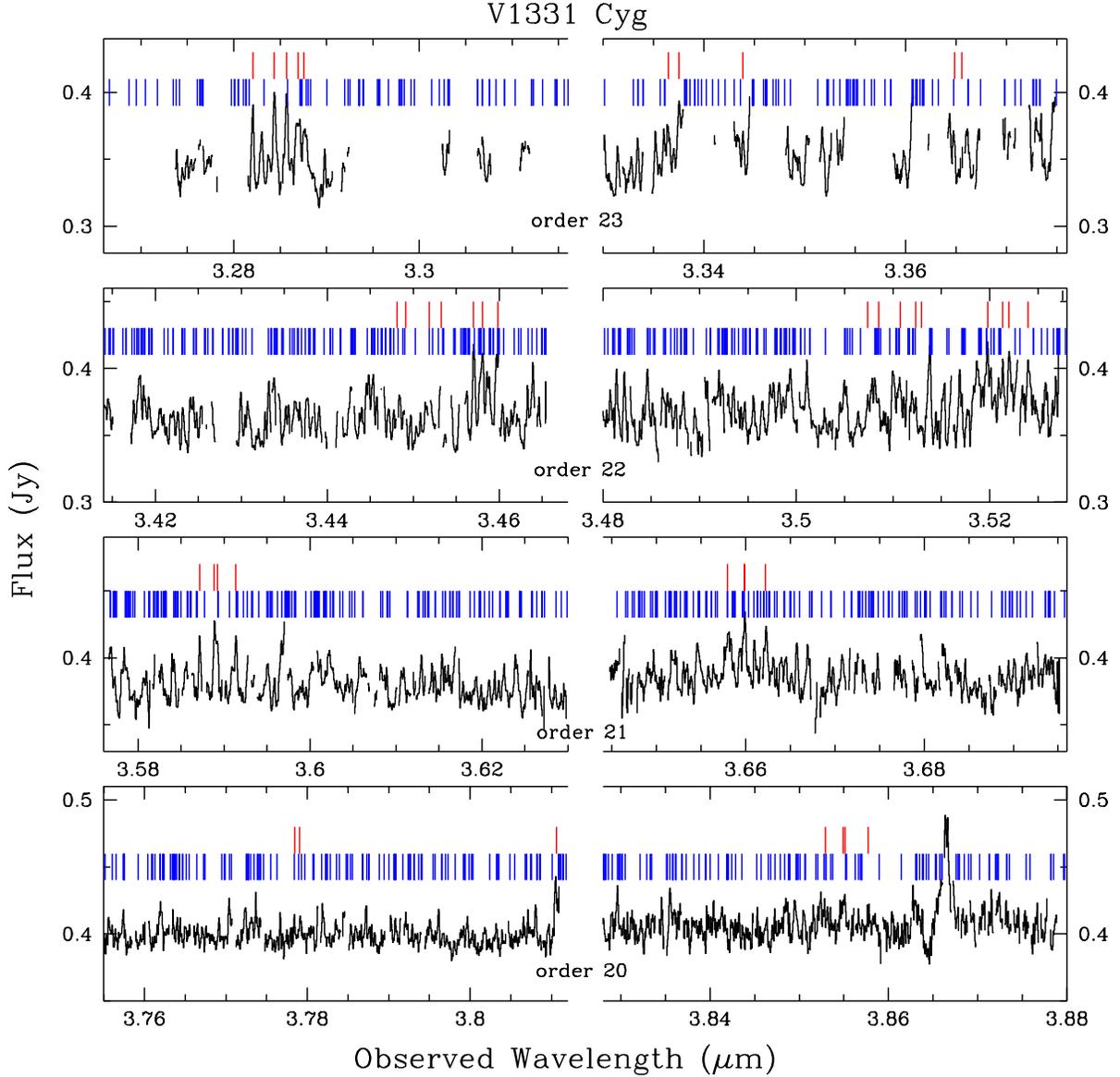}
\caption[]
{
High signal-to-noise (SNR $>$150), high resolution ($R\equiv
\lambda/\Delta \lambda=24,000$) spectra of V1331~Cyg spanning 4
$L$-band orders observed with NIRSPEC at two grating settings.
Emission lines of water (lower ticks, blue) and OH (upper ticks, red)
are evident throughout these $L$-band spectra.
Lines within 10\% of the strongest line in the BT2 list for each order were selected here.  
Wavelength regions
where the telluric transmission was poor (i.e., $\le$ 80\%) have been
excised from the plots.
}
\end{figure}

% Fig 2a Modeling water and OH
\begin{figure}
\figurenum{2a}
\plotone{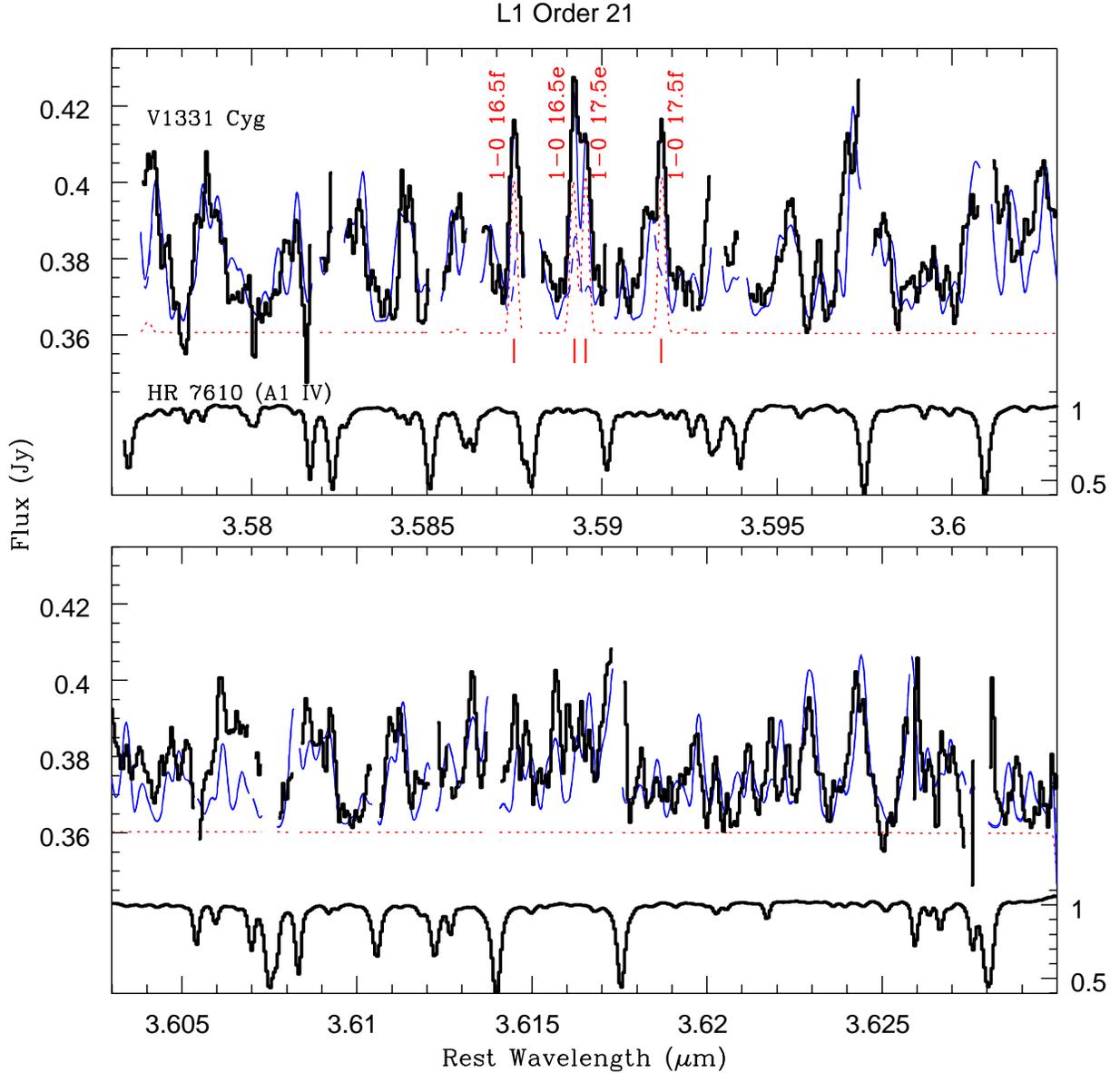}
\caption[Modeled Water Emission in V1331~Cyg:  L1ord21]
{ 
A rich infrared spectrum is observed in V1331~Cyg (upper thick
histogram, both panels) arising from emission by numerous water lines
blended with OH lines (vertical red ticks, labeled).  We model the
emission across multiple $L$-band orders (Figs. 2a--2h), and find a
good model fit (solid blue line) to the data by combining H$_2$O
(dashed blue line) and OH (dotted red line) model spectra synthesized
at $T=1500$\,K with different column densities ($2 \times 10^{21}$ and
$1 \times 10^{20}\persqcm$, respectively) and disk emitting areas
(5.5~$< R_{\rm H_2O}/\Rsun < 19$ and 5.5~$ < R_{\rm OH}/\Rsun < 28$,
respectively).  Telluric absorption is indicated by the spectrum of
the hot star, HR~7610 (lower thick histogram, both panels).
Wavelength regions in the data and models where the telluric
transmission is below 80\% are not plotted.  (A color version of the
figure is available in the online journal) 
}
\end{figure}

% Fig 2b Modeling water and OH
\begin{figure}
\figurenum{2b}
\plotone{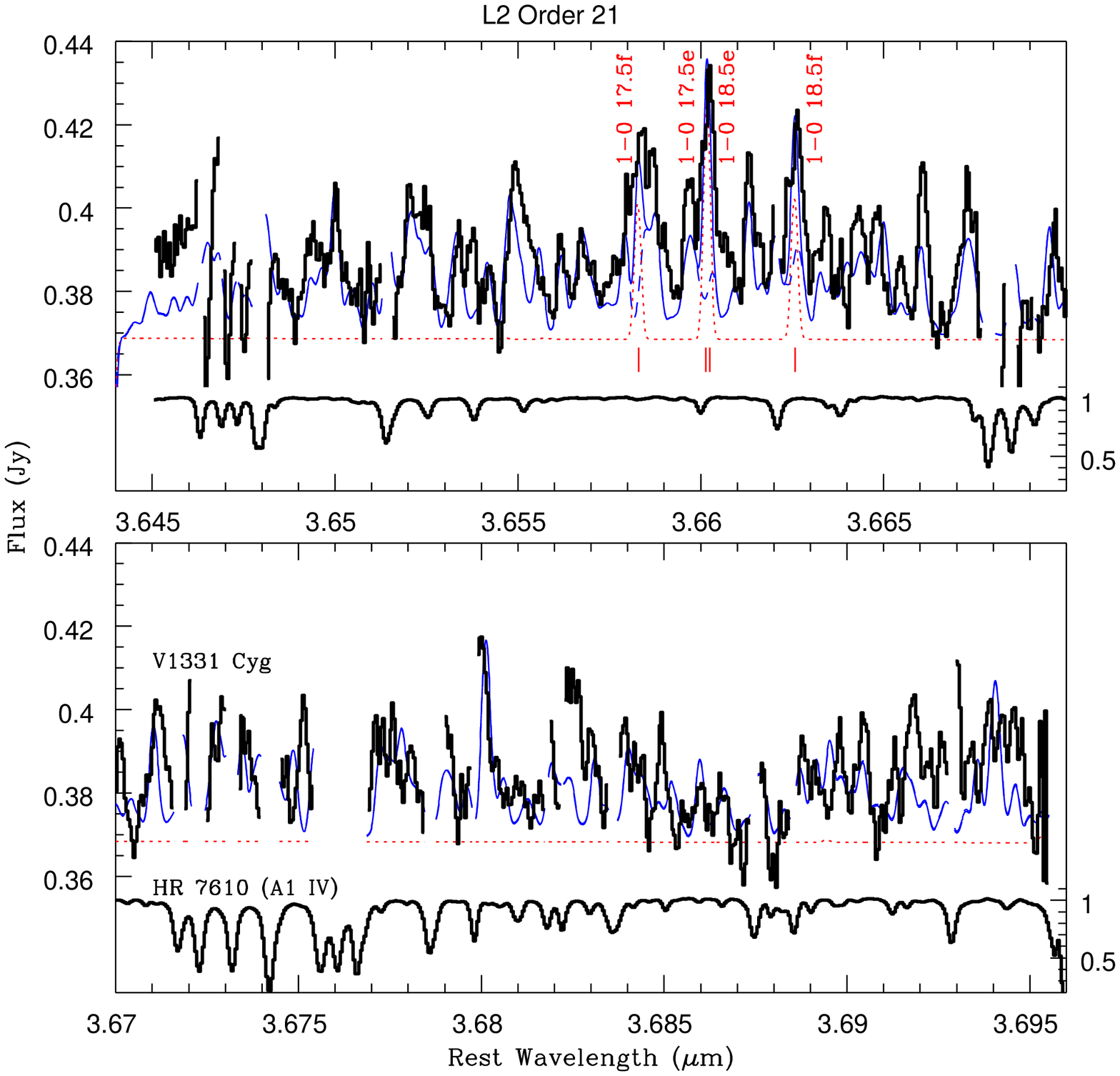}
\caption[Modeled Water Emission in V1331~Cyg:  L2ord21]
{
As in Fig. 2a but for the L2 order 21 wavelength region.
}
\end{figure}

% Fig 2c Modeling water and OH
\begin{figure}
\figurenum{2c}
\plotone{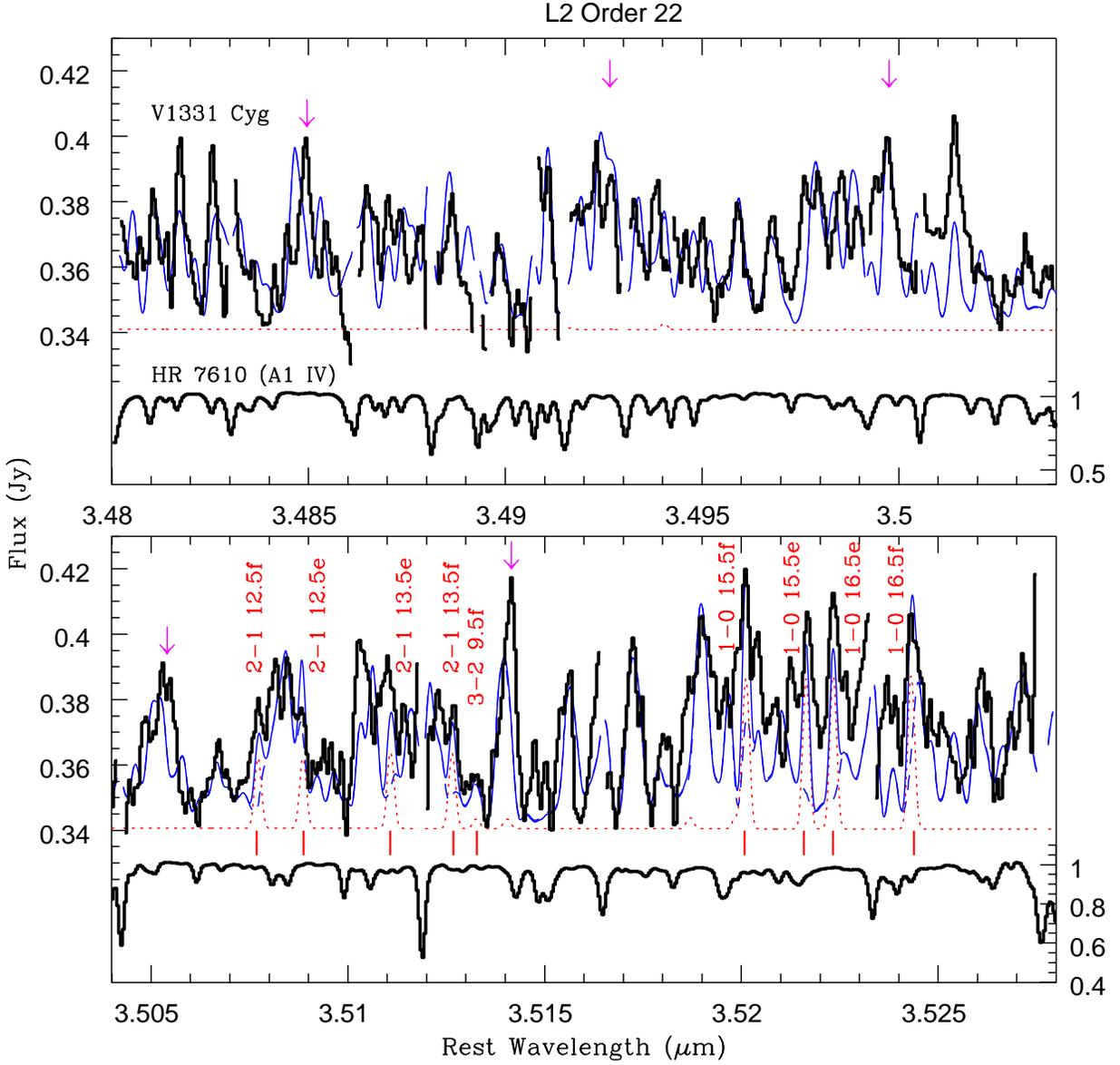}
\caption[Modeled Water Emission in V1331~Cyg:  L2ord22]
{ 
As in Fig. 2a but for the L2 order 22 wavelength region.  The OH
line emission (red ticks, lower panel) is strong and optically thick,
characterized by $\nu$~=~2--1 and 1--0 lines (see Table 1). While the
combined H$_2$O+OH emission model (solid blue) fits much of the
observed structure well, some wavelength offsets between the reference
model and the data are discernible (vertical arrows) along with
discrepancies in the strengths of some line groups.  The vertical
arrows illustrate some examples of water lines in this order that may
need empirical adjustments to their theoretically computed transition
wavelengths and/or oscillator strengths in the BT2 line list (see
$\S$3.2).  
}
\end{figure}

% Fig 2d Modeling water and OH
\begin{figure}
\figurenum{2d}
\plotone{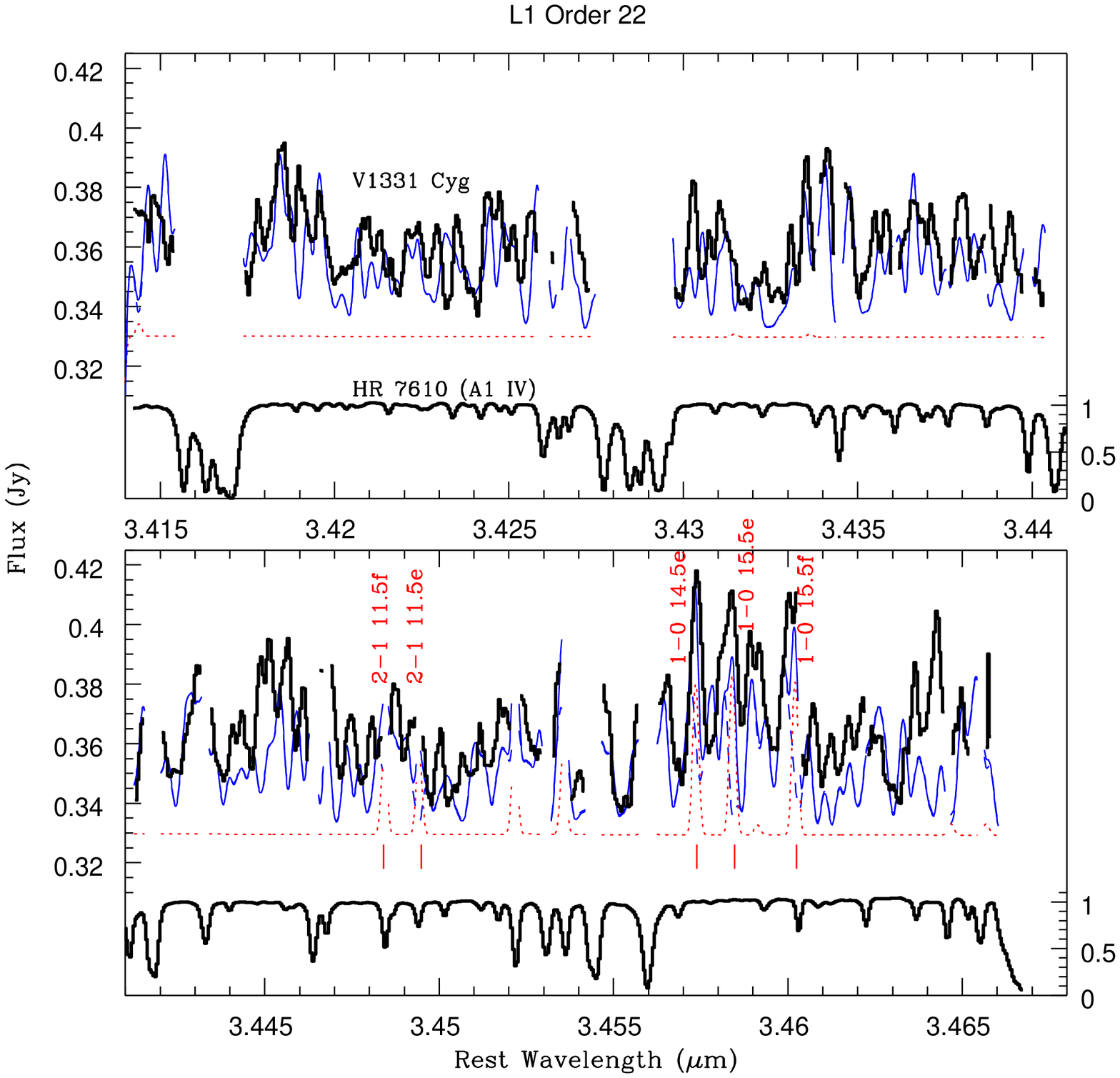}
\caption[Modeled Water Emission in V1331~Cyg:  L1ord22]
{
As in Fig. 2a but for the L1 order 22 wavelength region.
}
\end{figure}

% Fig 2e Modeling water and OH
\begin{figure}
\figurenum{2e}
\plotone{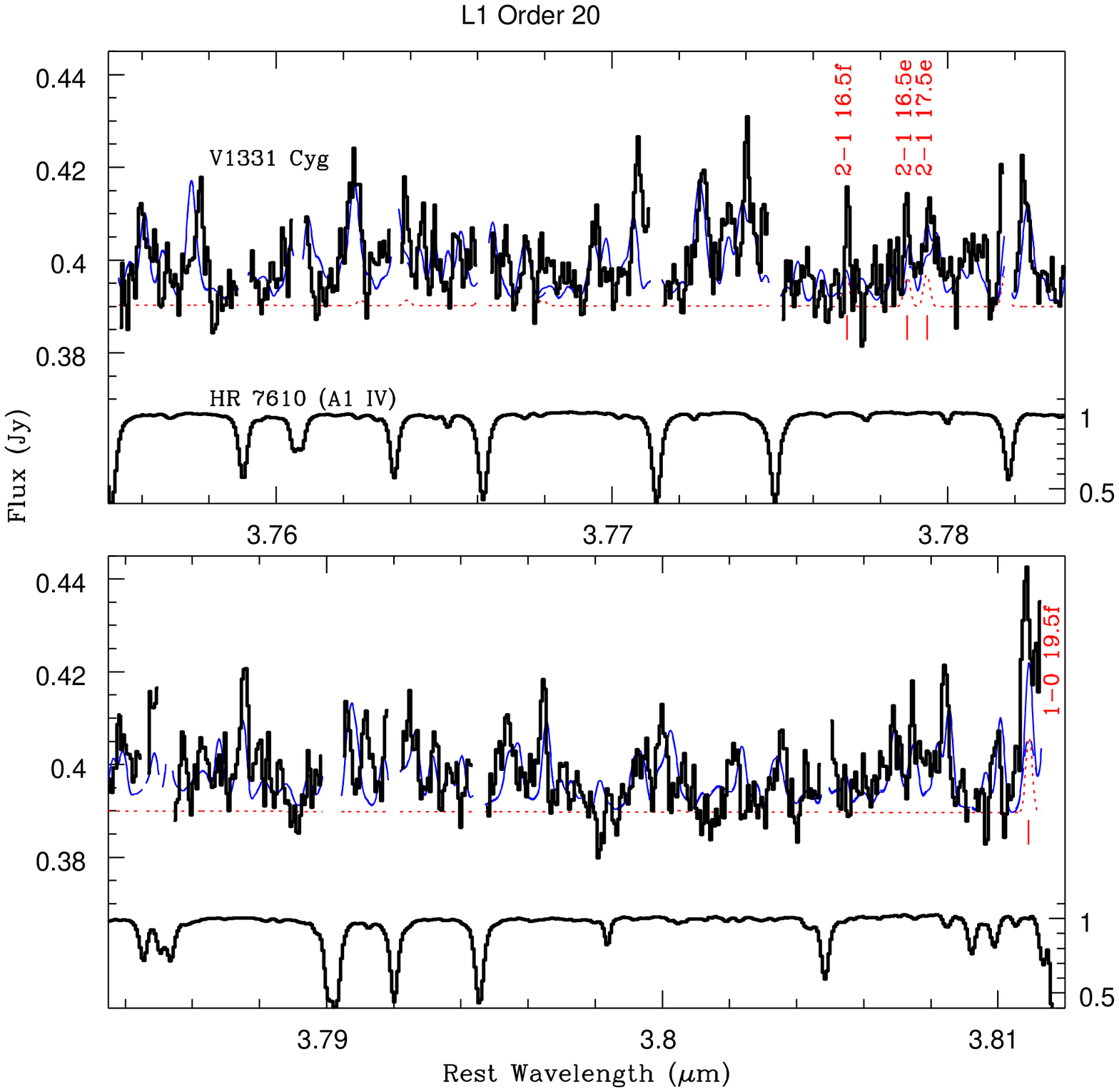}
\caption[Modeled Water Emission in V1331~Cyg:  L1ord20]
{
As in Fig. 2a but for the L1 order 20 wavelength region.
}
\end{figure}

% Fig 2f Modeling water and OH
\begin{figure}
\figurenum{2f}
\plotone{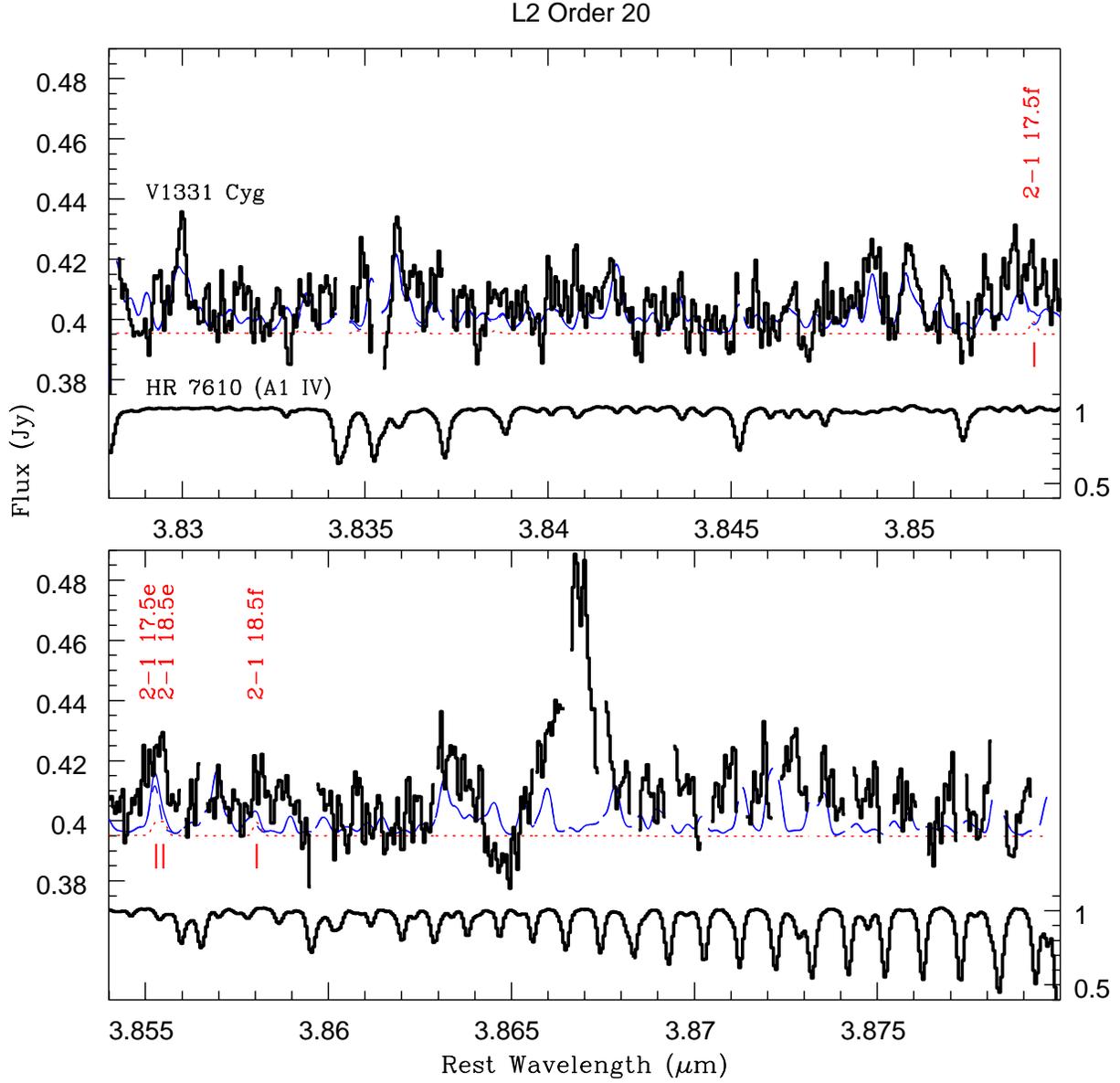}
\caption[Modeled Water Emission in V1331~Cyg:  L2ord20]
{ 
As in Fig. 2a but for the L2 order 20 wavelength region.  We observe
an unknown strong emission feature at 3.8667~$\mu$m, which could be
identified as a Mg I line ($\lambda_{\rm rest} = 3.86748~\mu$m) that
is blue shifted by 60 $\kms$ relative to the disk emission we detect,
perhaps as part of an outflowing wind.  
}
\end{figure}

% Fig 2g Modeling water and OH
\begin{figure}
\figurenum{2g}
\plotone{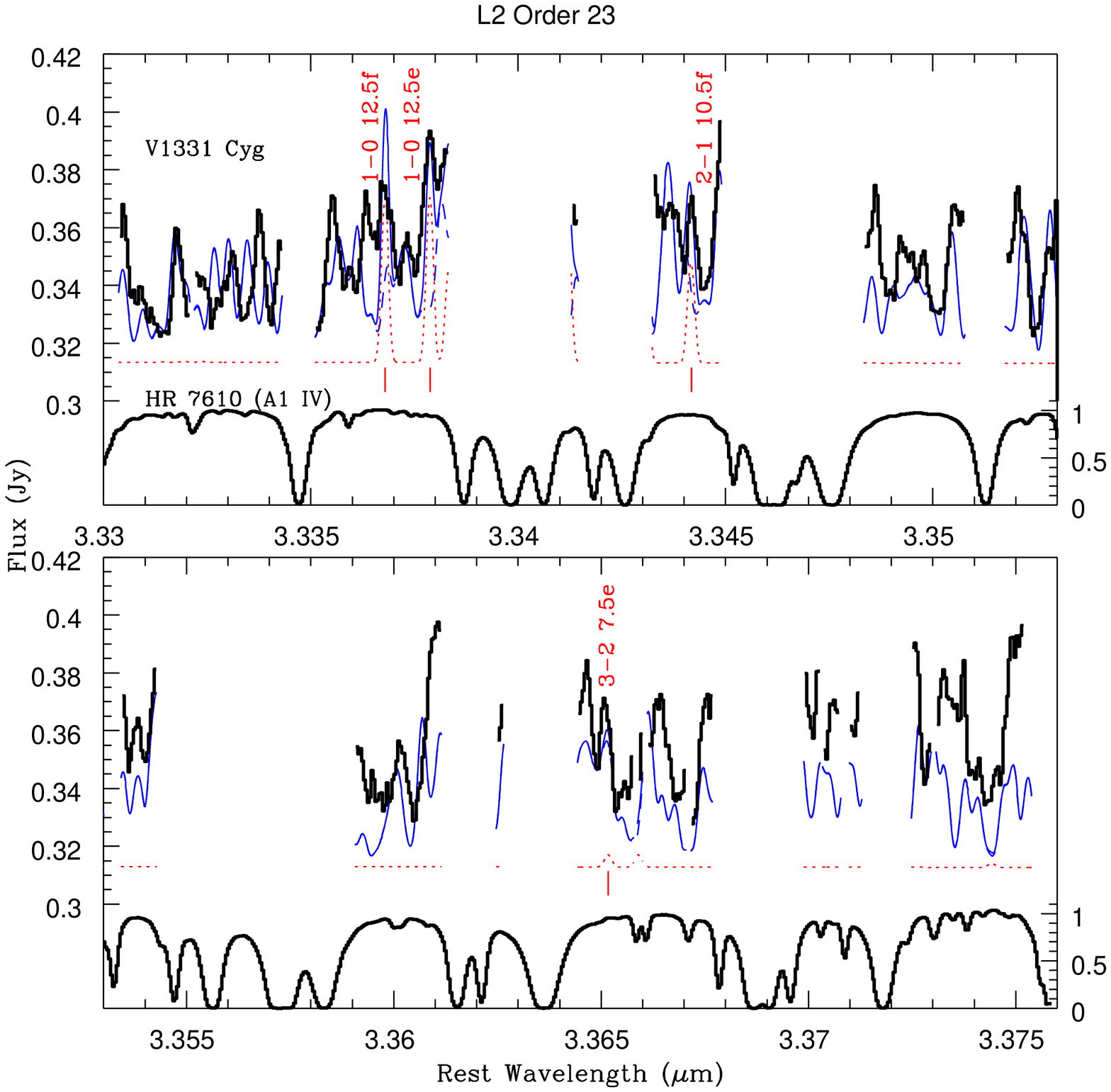}
\caption[Modeled Water Emission in V1331~Cyg:  L2ord23]
{
As in Fig. 2a but for the L2 order 23 wavelength region. 
}
\end{figure}

% Fig 2h Modeling water and OH
\begin{figure}
\figurenum{2h}
\plotone{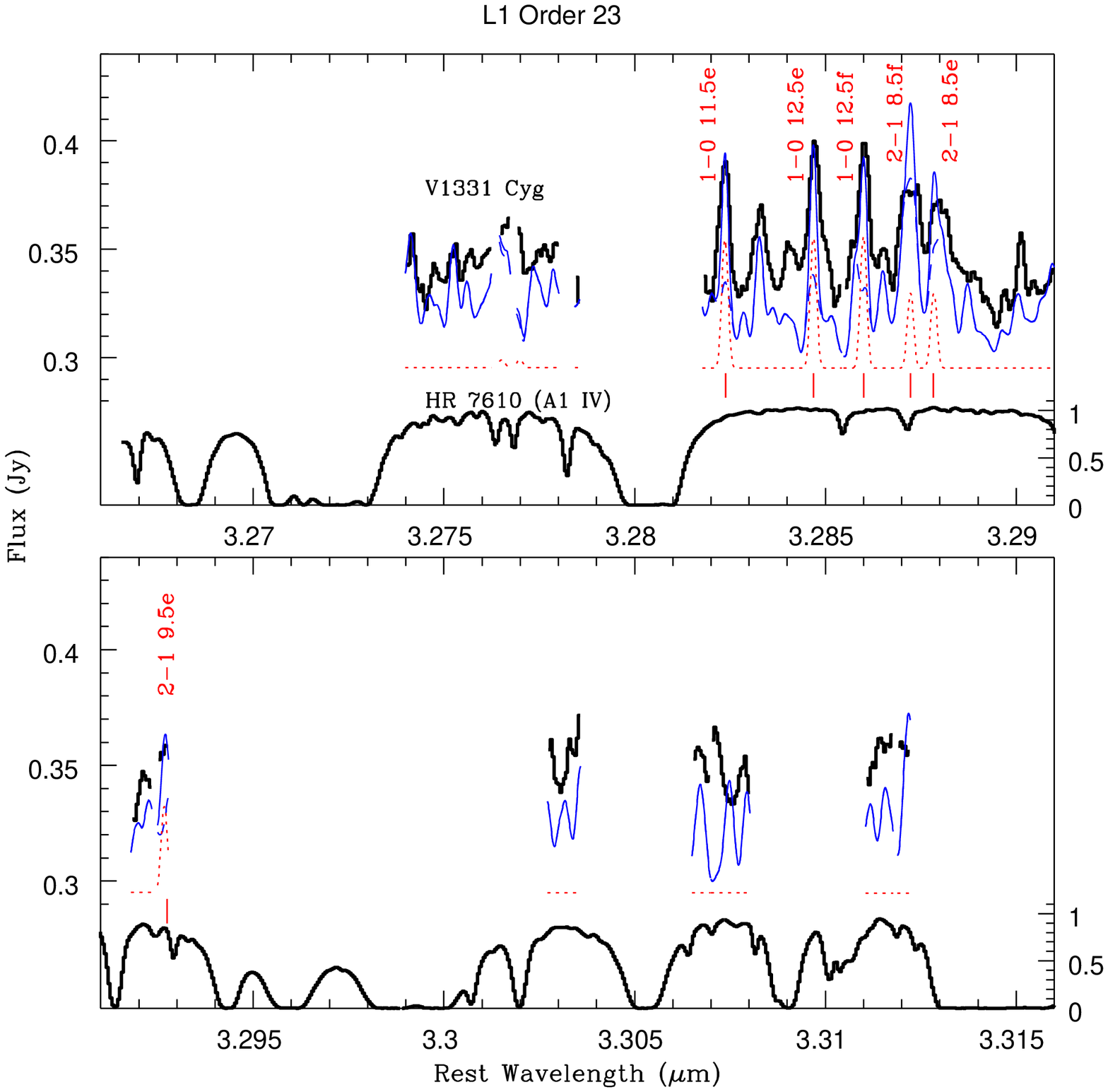}
\caption[Modeled Water Emission in V1331~Cyg:  L1ord23]
{
As in Fig. 2a but for the L1 order 23 wavelength region.
}
\end{figure}

% Fig 3a
\begin{figure}
\figurenum{3a}
\plotone{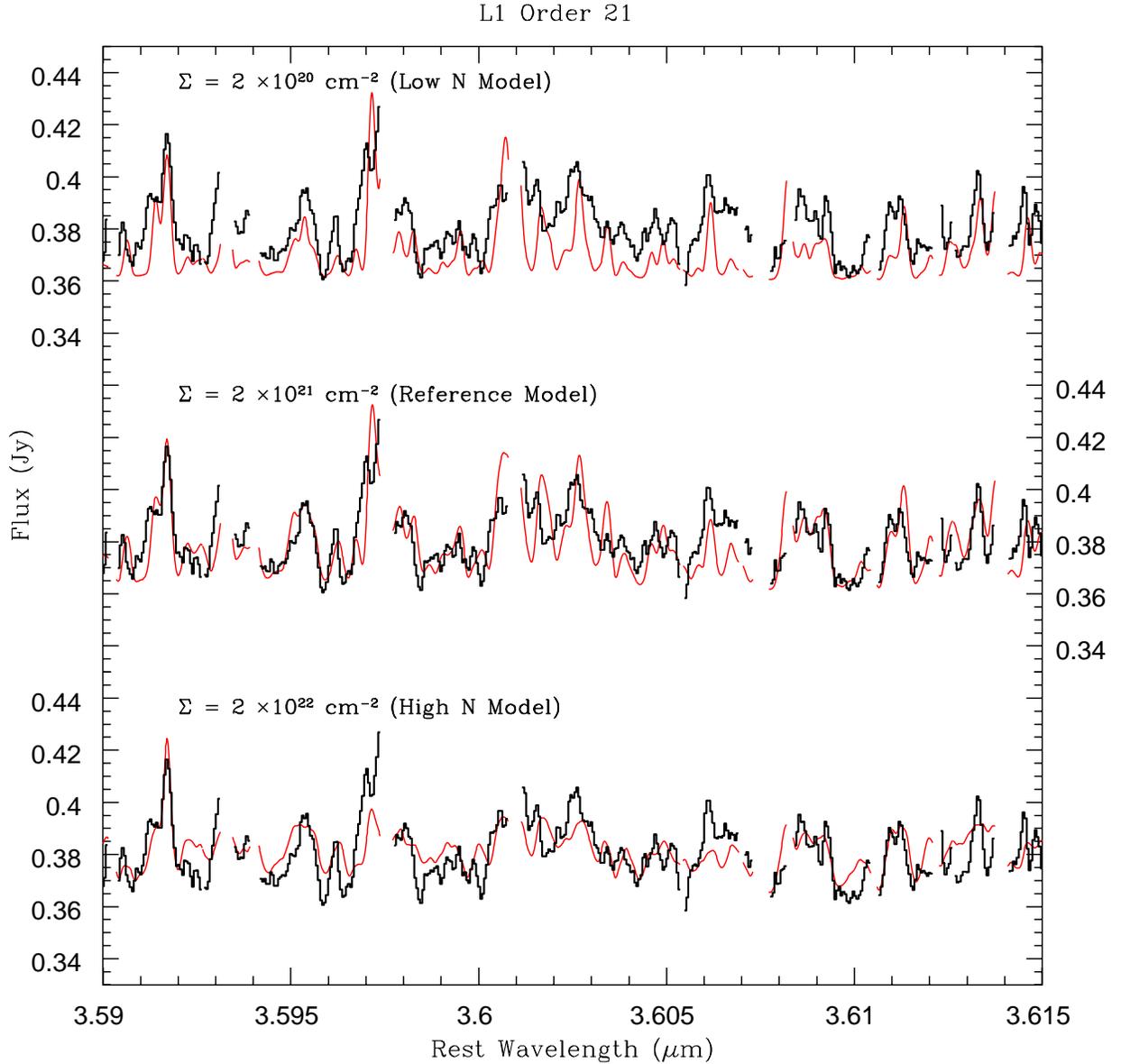}
\caption[]
{
Model fits of water emission to the data in a portion of the L1 order 21 region with different column density 
and emitting area values for a
fixed temperature ($T=1500\,$K). 
The emitting area was chosen to optimize the fit to the model flux level
in this order.  The low column density case ($N_{\rm H_2O}$=$2.0 \times 10^{20}
\persqcm, R_{\rm out} = 43\Rsun$, thin red line top panel) shows a poor fit
to the data (black histogram), having too much contrast between the strengths of the 
strong and weak lines, indicating that the lines are too optically
thin.  In the high column density case ($N_{\rm H_2O}$=$2.0 \times 10^{22}
\persqcm, R_{\rm out} = 12\Rsun$, thin red line lower panel), 
the fit to the data is also poor because the
strong and weak lines do not show enough contrast, since they are
much more optically thick.  The center panel shows a good fit to the
data ($N_{\rm H_2O}$=$2.0 \times 10^{21}
\persqcm, R_{\rm out} = 19\Rsun$) having a mix of optically 
thin and thick lines that reproduce the
strengths of both the strong and weak lines in this order.
}
\end{figure}

% Fig 3b
\begin{figure}
\figurenum{3b}
\plotone{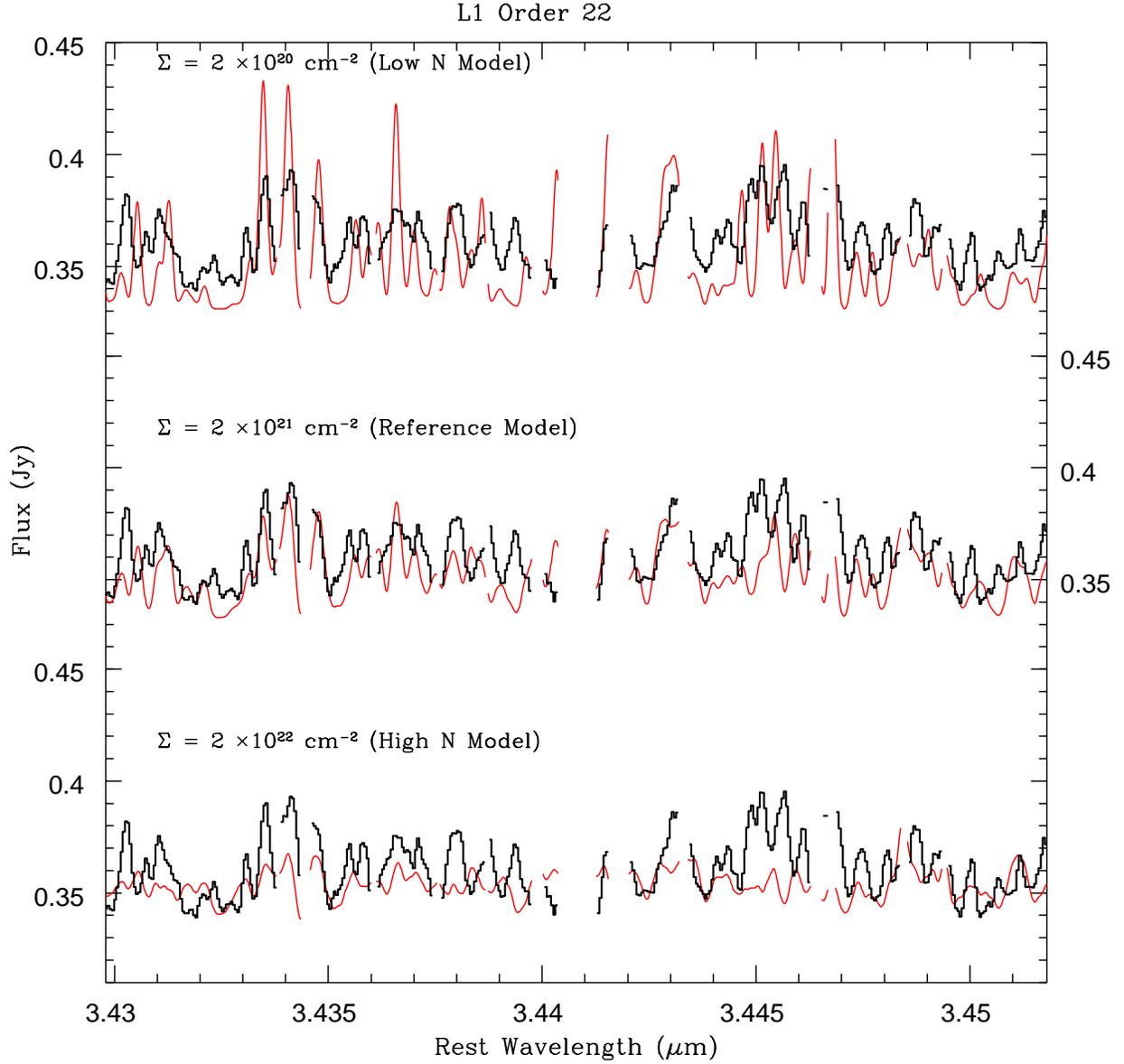}
\caption[]
{
As in Fig. 3a but for a portion of the L1 order 22 region.  The high and low column density models fit the
data more poorly here (compared to Fig. 3a) because the assumed emitting area is that chosen to optimize the fit to L1 order 21.
}
\end{figure}

% Fig 4a
\begin{figure}
\figurenum{4a}
\plotone{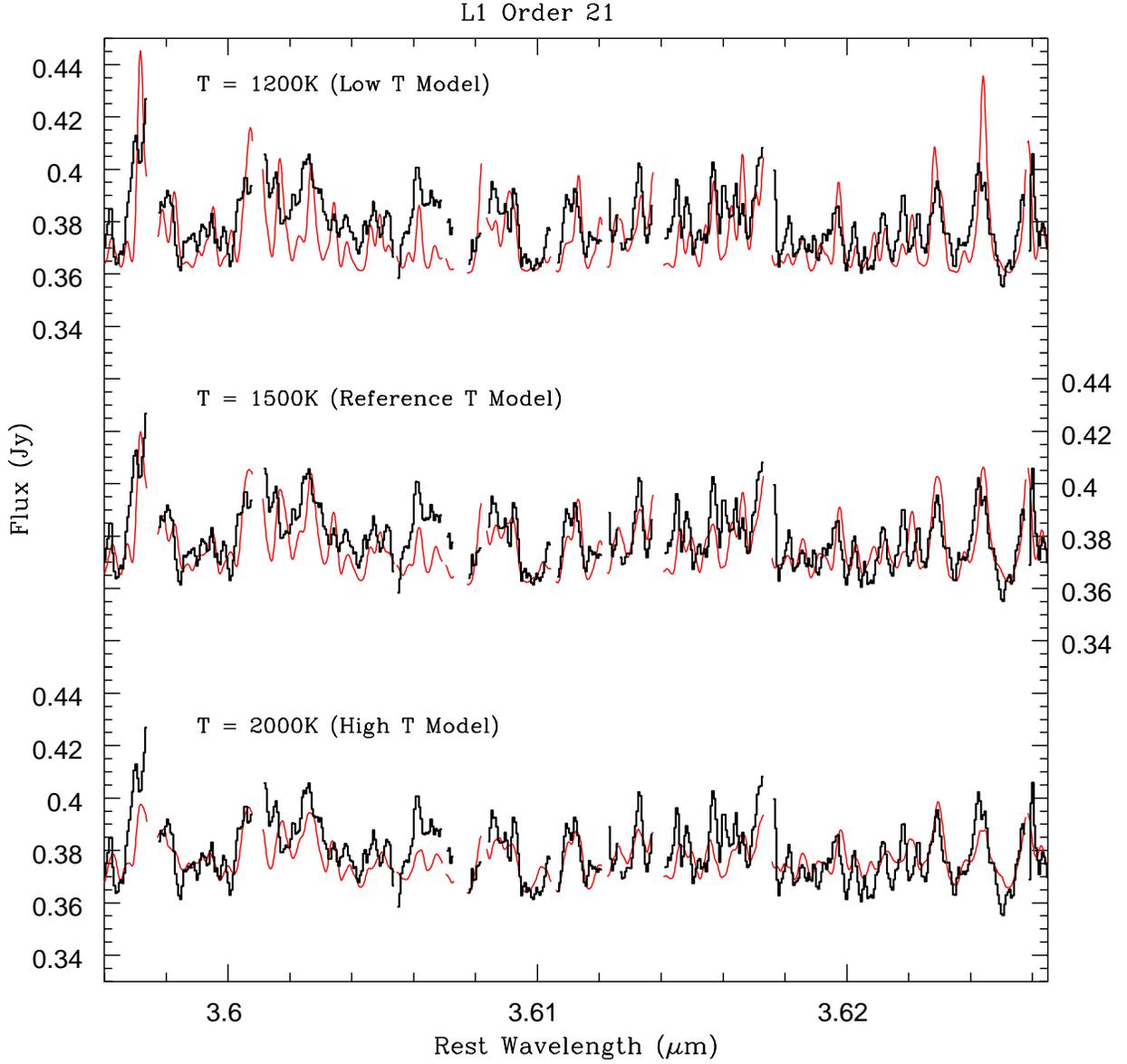}
\caption[]
{
Model fits of water emission to the data in a portion of the L1 order 21 region with temperature and emitting 
area values for a
fixed column density ($N=2 \times 10^{21} \persqcm$). The low temperature case ($T=1200K, R_{\rm out} = 
35\Rsun$, thin red line top panel ) shows a poor fit
to the data (thick black histogram), with the contrast between strengths of the 
strong and weak lines being too great.  In the high temperature case ($T=2000K, R_{\rm out} = 11\Rsun$, 
thin red line lower panel), 
the fit to the data is also poor because the
strong and weak lines do not show enough contrast.  The center panel shows a good fit to the
data ($T=1500K, R_{\rm out} = 19\Rsun$).
}
\end{figure}

% Fig 4b
\begin{figure}
\figurenum{4b}
\plotone{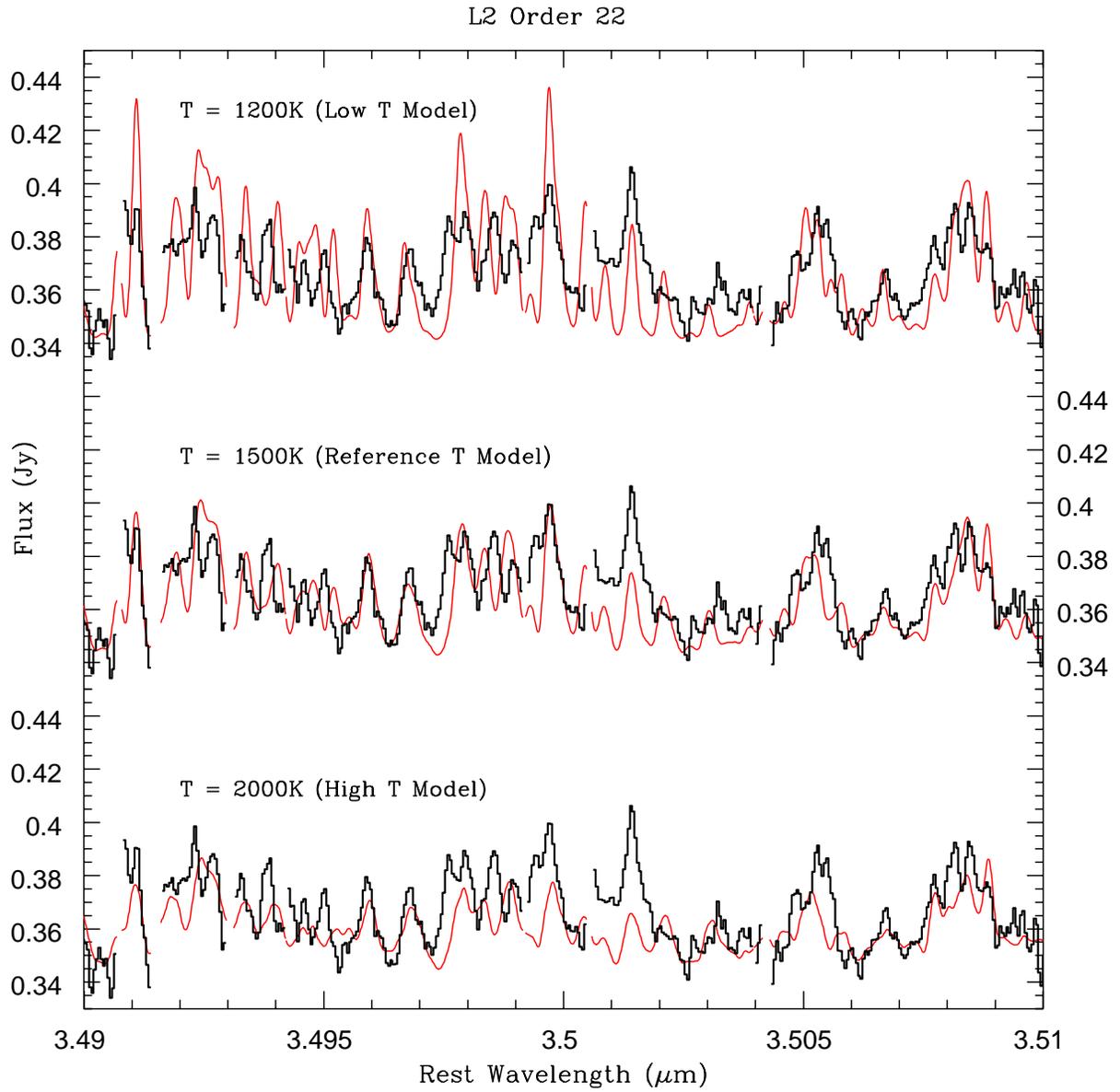}
\caption[]
{
As in Fig. 4a but for a portion of the L2 order 22 region.
}
\end{figure}

% Fig 5a  OH Model Ranges in Column Density

\begin{figure}
\figurenum{5a}
\plotone{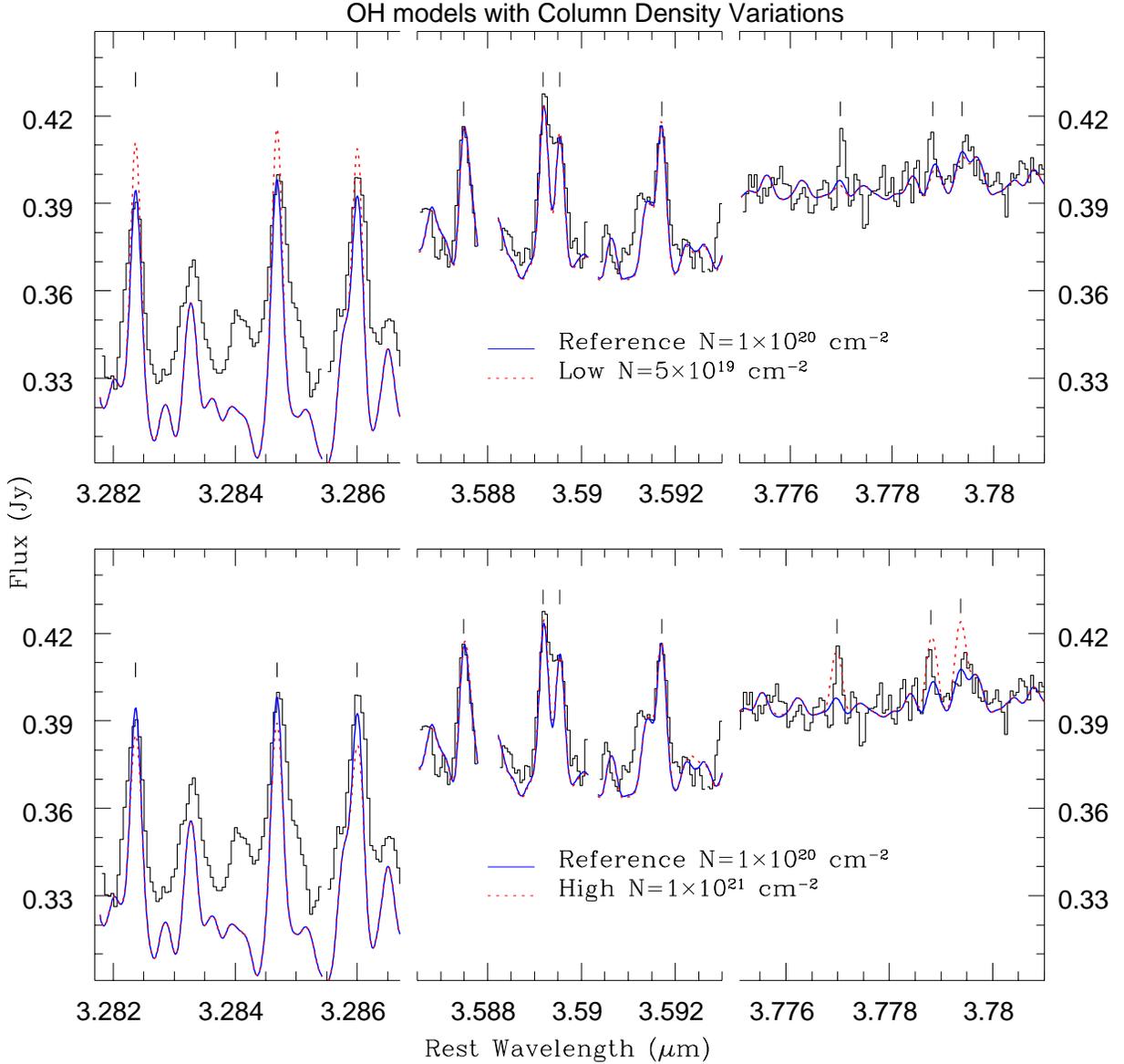}
\caption[]
{
OH model fits for higher and lower column densities than the reference OH model at a fixed temperature 
(T=1500\,K) compared with the observed V1331 Cyg spectrum (black histogram).  The top panel compares
the reference OH model (N=$1 \times 10^{20} \persqcm$, solid blue line) with a low column density model 
(N=$5 \times 10^{19} \persqcm$; dotted red line).  The low column density model overpredicts the flux of the 
lines at shorter wavelengths (3.282-3.287 $\mu$m).  The bottom panel compares the reference model (N=$1 
\times 10^{20} \persqcm$, solid blue line) with a high column density model (N=$1 \times 10^{21} \persqcm$, 
dotted red line).  The high column density model underpredicts the flux of 
the lines at shorter wavelengths (3.282-3.287 $\mu$m) and somewhat overpredicts the flux of the lines at 
longer wavelengths (3.776-3.780 $\mu$m).  The reference model fits all the OH lines (vertical ticks, see 
Table 1) adequately well across a range of optical depths.
}
\end{figure}

% Fig 5b  OH Model Ranges in Temperature

\begin{figure}
\figurenum{5b}
\plotone{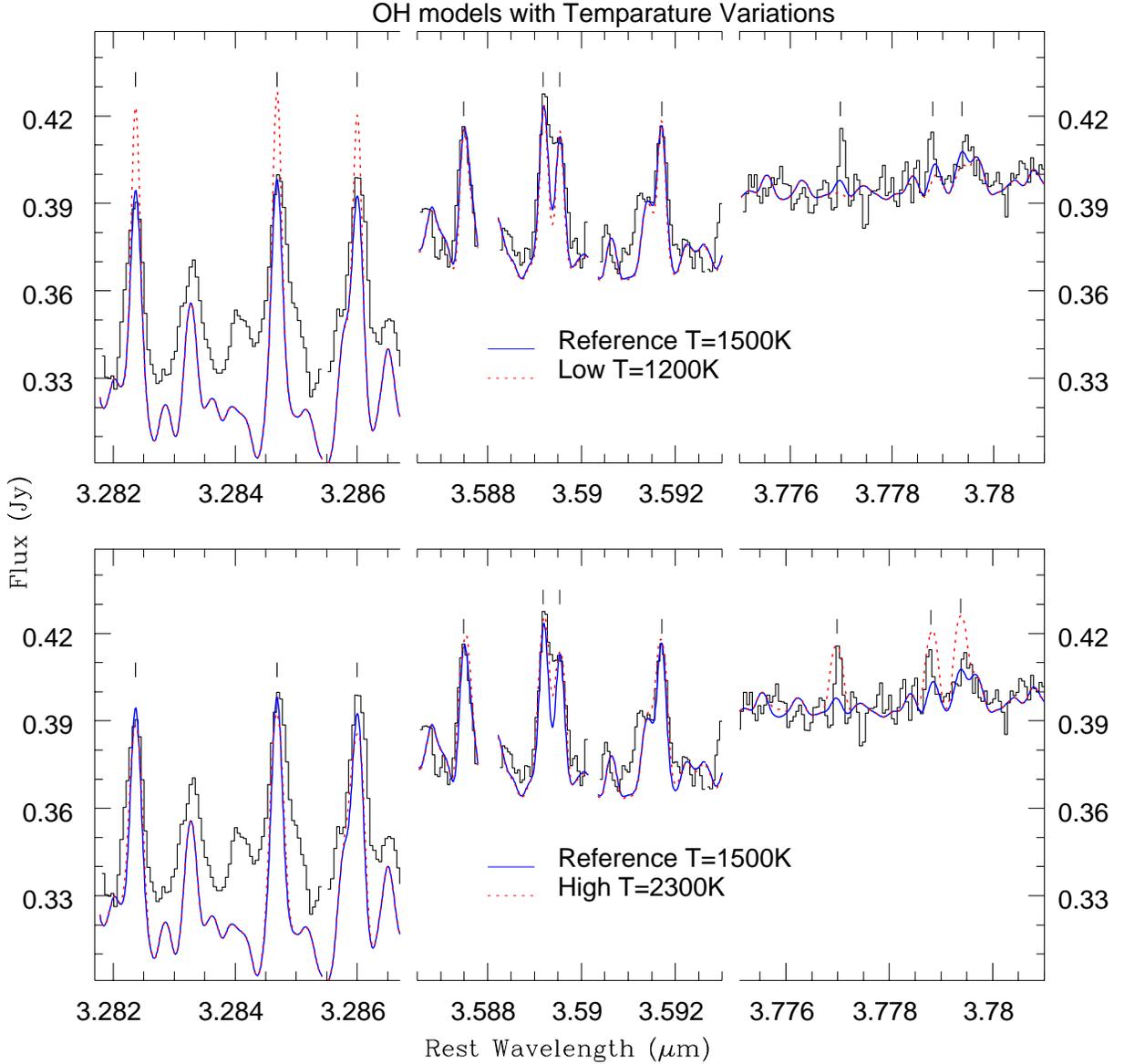}
\caption[]
{ 
OH model fits for higher and lower temperatures than the reference OH model at a fixed column density (N=
$1 \times 10^{20} \persqcm$) compared with the observed V1331 Cyg spectrum (black histogram).  The top 
panel compares the reference OH model (T=1500\,K, solid blue line) with a low temperature model 
(T=1200\,K; dotted red line).  The low temperature model overpredicts the flux of the low-J lines at shorter 
wavelengths (3.282-3.287 $\mu$m).  The bottom panel compares the reference model (T=1500\,K, solid 
blue line) with a high temperature model (T=2300\,K, dotted red line).  The high temperature model 
overpredicts the flux of 
the high-J lines at longer wavelengths (3.776-3.780 $\mu$m).  The low temperature model puts too much 
energy into the low-J lines (shorter wavelengths, top panel), and the high temperature model puts too much 
energy into the high-J lines (longer wavelengths, bottom panel). The reference model fits all the OH lines 
(vertical ticks, see Table 1) adequately well across a range of excitation levels.
}
\end{figure}

% Fig 6 BT vs. Hitran plot
\begin{figure}
\figurenum{6}
\plotone{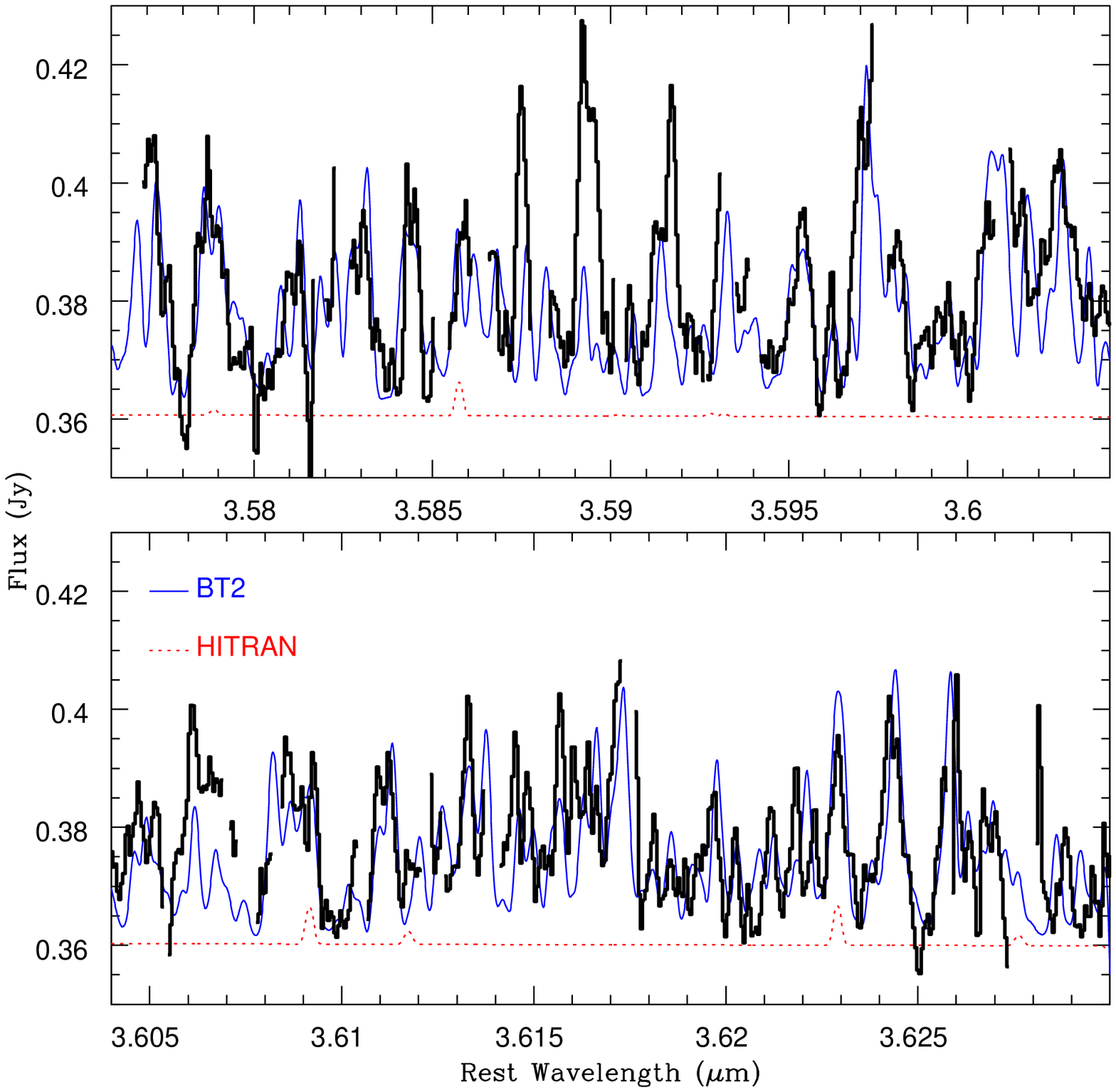}
\caption[Energies of BT2 lines]
{ 
A comparison of synthetic spectra at the same temperature
($T=1500\,$K), H$_2$O column density ($2 \times 10^{21} \persqcm$),
and emitting area (5.5~$< R_{\rm H_2O}/\Rsun < 19$) as was found for
the reference model to V1331~Cyg, but generated using different water
line lists.  The model using the BT2 water list (solid blue line) fits
most of the observed structure in V1331~Cyg (black histogram), while
the model from the HITRAN water line list (dotted red line) does a
poor job fitting the data (thick black histogram), with very few
emission features present arising from a depressed continuum level.
}
\end{figure}

% Fig 7 CO and H2O fit
\begin{figure}
\figurenum{7}
\plotone{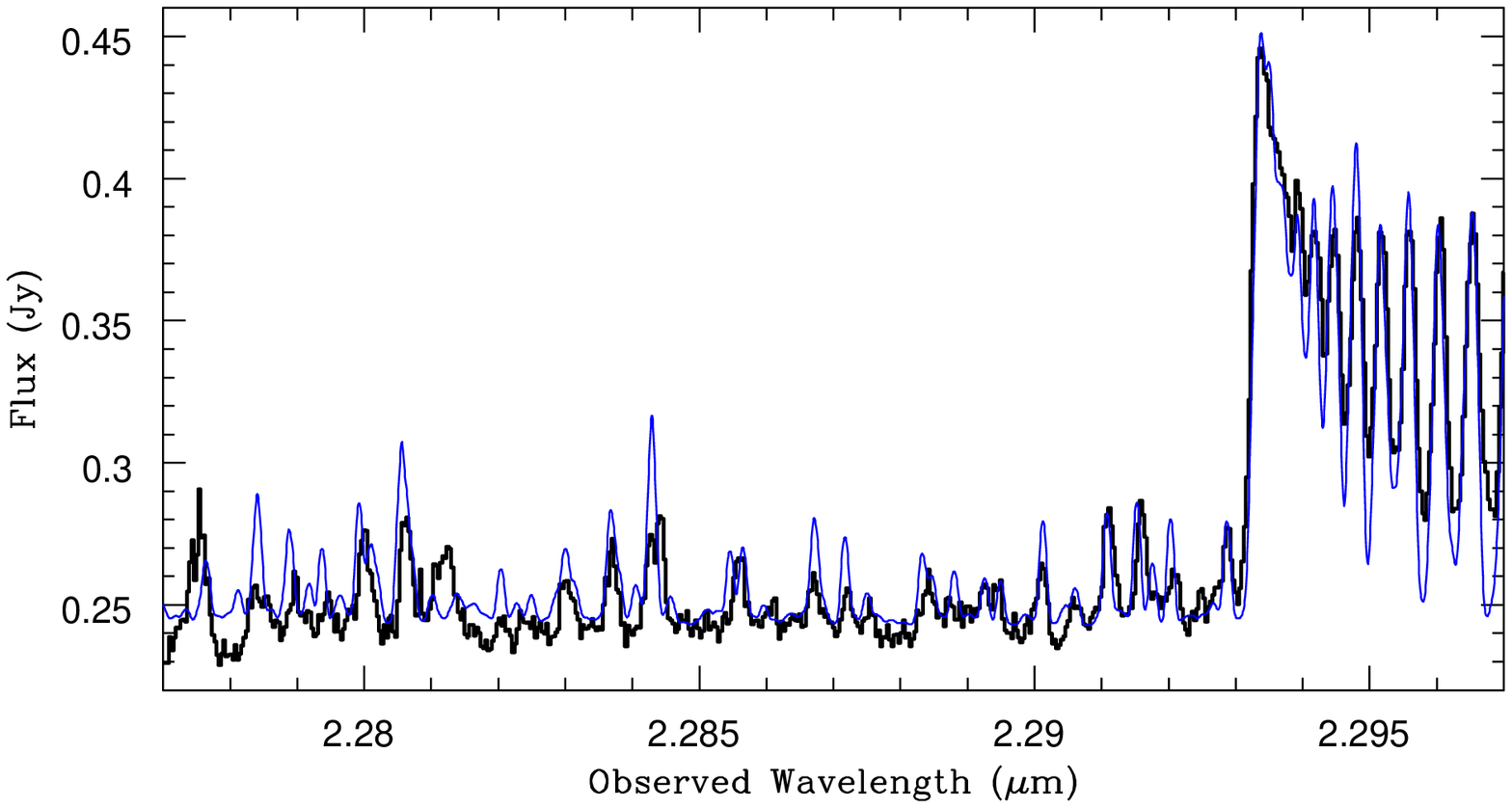}
\caption[H2O + CO fit]
{ 
A fit to the $K$-band spectrum of V1331~Cyg observed in an earlier
epoch (thick histogram, Najita et al. 2009), showing strong water
emission blueward of the overtone CO emission bandhead.  The model
fit, a combination of water and CO emission (thin blue line), does a
good job fitting the observed structure, highlighting the utility of
the BT2 list in fitting warm $K$-band water emission.  The fit
parameters for the water emission in the $K$-band are the same as for
the water in the $L$-band (i.e., Fig. 2), except for having an
emitting area that is 1.6 times larger, which could be ascribed to
time variable emission and/or uncertainties in the flux calibration
between the two observational epochs.  The CO model has a temperature
of 1800~K, a column density of $6 \times 10^{21} \persqcm$, and an
emitting area extending from $\Rin = 2.3 \Rsun$ to $\Rout = 8\Rin$
(see Table 2).
}
\end{figure}

\clearpage
\oddsidemargin=-1cm
\tabletypesize{\scriptsize}

% TABLES

%% 
%% Begining of file `tab1.tex'

%  Table of OH line properties

\begin{deluxetable}{cccccc}
\tablecolumns{4}
\tablewidth{0 pt}
\tablecaption{Physical properties of OH lines detected in our observations}
\tablehead{
\colhead{$L$-band Order}   				& 
\colhead{Transition $\lambda$($\rm \AA$)}   		& 
\colhead{Line Transition Details} 			&
\colhead{Model Line Opacity$\tablenotemark{a}$} 		
}
\startdata

L1 order 23 & 32823.671 & $\nu$=1-0; J=11.5e & 22.9 \\
		   & 32846.868 & $\nu$=1-0; J=12.5e & 26.0 \\
		   & 32859.990 & $\nu$=1-0; J=12.5f & 25.8 \\
		   & 32872.299 & $\nu$=2-1; J=8.5f  & 2.5  \\
		   & 32878.437 & $\nu$=2-1; J=8.5e & 2.5 \\
		   & 32926.693 & $\nu$=2-1; J=9.5e & 2.9 \\

L2 order 23 & 33367.886 & $\nu$=1-0; J=12.5f &17.2\\
 		  & 33378.685 & $\nu$=1-0; J=12.5e & 17.0 \\
		  & 33441.684 & $\nu$=2-1; J=10.5f & 2.5  \\
		  & 33651.661 & $\nu$=3-2; J=7.5e & 0.16 \\

L1 order22 & 34484.234 & $\nu$=2-1; J=11.5f & 1.4 \\
 		 & 34494.249 & $\nu$=2-1; J=11.5e & 1.4\\
 		 & 34573.554 & $\nu$=1-0; J=14.5e &  8.4\\
 		 & 34584.036 & $\nu$=1-0; J=15.5e & 9.3 \\
 		 & 34602.034 & $\nu$=1-0; J=15.5f & 9.2 \\

L2 order 22 & 35077.159 & $\nu$=2-1; J=12.5f &1.1 \\
 		  & 35088.615 & $\nu$=2-1; J=12.5e & 1.1\\
 		  & 35110.880 & $\nu$=2-1; J=13.5e & 1.2\\
 		  & 35126.505 & $\nu$=2-1; J=13.5f & 1.2\\
 		  & 35132.676 & $\nu$=3-2; J= 9.5f & 0.11\\
 		  & 35201.132 & $\nu$=1-0; J=15.5f & 5.6\\
 		  & 35216.445 & $\nu$=1-0; J=15.5e & 5.5\\
 		  & 35223.349 & $\nu$=1-0; J=16.5e & 6.1\\
 		  & 35243.170 & $\nu$=1-0; J=16.5f & 6.0\\

L1 order 21 & 35874.842 & $\nu$=1-0; J=16.5f & 3.6\\
 		  & 35891.844 & $\nu$=1-0; J=16.5e & 3.5\\
 		  & 35895.358 & $\nu$=1-0; J=17.5e & 3.9\\
 		  & 35917.113 & $\nu$=1-0; J=17.5f & 3.8\\
 
L2 order 21 & 36582.769 & $\nu$=1-0; J=17.5f & 2.2 \\
 		  & 36601.565 & $\nu$=1-0; J=17.5e & 2.2\\
 		  & 36601.815 & $\nu$=1-0; J=18.5e & 2.4 \\
 		  & 36625.624 & $\nu$=1-0; J=18.5f & 2.4 \\

L1 order 20 & 37788.009 & $\nu$=2-1; J=16.5e & 0.24 \\
 		  & 37793.784 & $\nu$=2-1; J=17.5e & 0.27 \\
 		  & 38109.260 & $\nu$=1-0; J=19.5f &  0.75 \\
 
L2 order 20  & 38532.612 & $\nu$=2-1; J=17.5f & 0.16 \\
 		   & 38552.724 & $\nu$=2-1; J=17.5e  & 0.15 \\
 		   & 38554.918 & $\nu$=2-1; J=18.5e & 0.17 \\
 		   & 38580.509 & $\nu$=2-1; J=18.5f  & 0.17 \\

\enddata

\tablenotetext{a}{~OH reference model: $T=1500\,$K, $N=1.0 \times 10^{20} \persqcm$}

\end{deluxetable}

%% 
%% Begining of file `tab2.tex'

%  Table of OH line properties

\begin{deluxetable}{cccccccc}
\tablecolumns{8}
\tablewidth{0 pt}
\tablecaption{Comparison of molecular gas parameters derived for V1331~Cyg}
\tablehead{
\colhead{Obs. Date}	&
\colhead{$\lambda$}	&
\colhead{Molecule}		&
\colhead{$T$}			& 
\colhead{$N$}	& 
\colhead{Emitting Radii}		&
\colhead{$v_{\rm turb}$}		&
\colhead{$N_{\rm Mol}/N_{\rm H_2O} $} \\
\colhead{(UT)}	&
\colhead{}	&
\colhead{}		&
\colhead{(K)}			& 
\colhead{($\times10^{21}\persqcm$)}	& 
\colhead{($\Rsun$)}		&
\colhead{($\kms$)}		&
\colhead{} \\
}

\startdata

1999 July 3  & K  &  CO		&  1800		& 6		& 2.3 -- 18	& 4    &  3--10\\
1999 July 3  & K  &  H$_2$O	&  1500		& 0.6		& 5.5 -- 26	& 4    &  1\\
1999 July 3  & K  &  H$_2$O	&  1500		& 2		& 5.5 -- 26	& 0    &  1\\
2001 July 11& L  &  H$_2$O	&  1500		& 2		& 5.5 -- 19	& 0    &  1\\
2001 July 11& L  &  OH		&  1500		& 0.1		& 5.5 -- 28	& 0    &  0.05\\

\enddata
\end{deluxetable}


\begin{thebibliography}{}


\bibitem[Ag{\'u}ndez et al.(2008)]{agundez2008} Ag{\'u}ndez, M., Cernicharo, J., \& Goicoechea, J.~R.\ 2008, \aap, 483, 831

\bibitem[Alves et al.(1998)]{alves1998} Alves, J., Lada, C.~J., Lada, E.~A., Kenyon, S.~J., \& Phelps, R.\ 1998, \apj, 506, 292 

\bibitem[Balbus \& Hawley(1991)]{balbus1991} Balbus, S.~A., \& Hawley, J.~F.\ 1991, \apj, 376, 214

\bibitem[Bethell \& Bergin(2009)]{bethell2009} Bethell, T., \& Bergin, E.\ 2009, Science, 326, 1675 

\bibitem[Barber et al.(2006)]{barber2006} Barber, R.~J., Tennyson, J., Harris, G.~J., \& Tolchenov, R.~N.\ 2006, \mnras, 368, 1087 (BT2)

\bibitem[Branham(1982)]{branham1982} Branham, R.~L., Jr.\ 1982, \aj, 87, 928 

\bibitem[Carr(1989)]{carr1989} Carr, J.~S.\ 1989, \apj, 345, 522 

\bibitem[Carr et al.(2004)]{carr2004} Carr, J.~S., Tokunaga, A.~T., \& Najita, J.\ 2004, \apj, 603, 213

\bibitem[Carr \& Najita(2008)]{carr2008} Carr, J.~S., \& Najita, J.~R.\ 2008, Science, 319, 1504

\bibitem[Carr \& Najita(2011)]{carr2011} Carr, J.~S., \& Najita, J.~R.\ 2011, accepted

\bibitem[Chavarr\'{i}a(1981)]{chavarria1981} Chavarr\'{i}a, C.\ 1981, \aap, 101, 105 

\bibitem[Eisner et al.(2007)]{eisner2007} Eisner, J.~A., Hillenbrand, L.~A., White, R.~J., Bloom, J.~S., Akeson, R.~L., \& Blake, C.~H.\ 2007, \apj, 669, 1072

\bibitem[Fedele et al.(2011)]{fedele2011} Fedele, D., Pascucci, I., Brittain, S., Kamp, I., Woitke, P., Williams, J.~P., Dent, W.~R.~F., \& Thi, W.~-.\ 2011, arXiv:1103.6039 

\bibitem[Glassgold et al.(2004)]{glassgold2004} Glassgold, A.~E., Najita, J., \& Igea, J.\ 2004, \apj, 615, 972

\bibitem[Glassgold et al.(2009)]{glassgold2009} Glassgold, A.~E.,  Meijerink, R., \& Najita, J.~R.\ 2009, \apj,701, 142 

\bibitem[Hamann \& Persson(1992)]{hamann1992} Hamann, F., \& Persson, S.~E.\ 1992, \apj, 394, 628 

\bibitem[Herbig \& Dahm(2006)]{herbig2006} Herbig, G.~H., \& Dahm, S.~E.\ 2006, \aj, 131, 1530 

\bibitem[Kamp et al.(2005)]{kamp2005} Kamp, I., Jonkheid, B., Augereau, J.-C., \& van Dishoeck, E.\ 2005, Nearby Resolved Debris Disks, 15 

\bibitem[Knez et al.(2007)]{knez2007} Knez, C., et al.\ 2007, \baas, 38, 812 

\bibitem[Kuhi(1964)]{kuhi1964} Kuhi, L.~V.\ 1964, Ph.D.~Thesis

\bibitem[Levreault(1988)]{levreault1988} Levreault, R.~M.\ 1988, \apj, 330, 897 

\bibitem[Mandell et al.(2008)]{mandell2008} Mandell, A.~M., Mumma, M.~J., Blake, G.~A., Bonev, B.~P., Villanueva, G.~L., \& Salyk, C.\ 2008, \apjl, 681, L25

\bibitem[McLean et al.(1998)]{mclean1998} McLean, I. S. et al. 1998, \procspie, 3354, 566

\bibitem[Massey et al.(1992)]{massey1992} Massey, P., Valdes, F., Barnes, J. 1992 A Users's Guide to Reducing Slit Spectra with \textit{IRAF}, National Optical Astronomy Observatory

\bibitem[Massey et al.(1997)]{massey1997} Massey, P.\ 1997 A User's Guide to CCD reductions with \textit{IRAF}, National Optical Astronomy Observatory

\bibitem[McMuldroch(1993)]{mcmuldroch1993} McMuldroch, S., Sargent, A.~I., \& Blake, G.~A.\ 1993, \aj, 106, 2477 

\bibitem[Najita et al.(1996)]{najita1996} Najita, J., Carr, J.~S., Glassgold, A.~E., Shu, F.~H., \& Tokunaga, A.~T. 1996, \apj, 462, 919

\bibitem[Najita et al.(2000)]{najita2000} Najita, J.~R., Edwards, S., Basri, G., \& Carr, J.\ 2000, Protostars and Planets IV, 457

\bibitem[Najita et al.(2003)]{najita2003} Najita, J., Carr, J.~S., \& Mathieu, R.~D.\ 2003, \apj, 589, 931

\bibitem[Najita et al.(2007)]{najita2007} Najita, J.~R., Carr, J.~S., Glassgold, A.~E., \& Valenti, J.~A.\ 2007, Protostars and Planets V, 507

\bibitem[Najita et al.(2009)]{najita2009} Najita, J.~R., Doppmann, G.~W., Carr, J.~S., Graham, J.~R., \& Eisner, J.~A.\ 2009, \apj, 691, 738

\bibitem[Partridge \& Schwenke(1997)]{partridge1997} Partridge,~H., \& Schwenke,~D.\ 1997, J. Chem. Phys., 106, 4618 (PS)

\bibitem[Pontoppidan et al.(2005)]{pontoppidan2005} Pontoppidan, K.~M., Dullemond, C.~P., van Dishoeck, E.~F., Blake, G.~A., Boogert, A.~C.~A., Evans, N.~J., II, Kessler-Silacci, J.~E., \& Lahuis, F.\ 2005, \apj, 622, 463 

\bibitem[Pontoppidan et al.(2008)]{pontoppidan2008} Pontoppidan, K.~M., et al.\ 2008, \apj, 678, 1005

\bibitem[Pontoppidan et al.(2010a)]{pontoppidan2010a} Pontoppidan, K.~M., Salyk, C., Blake, G.~A., Meijerink, R., Carr, J.~S., \& Najita, J.\ 2010, \apj, 720, 887

\bibitem[Pontoppidan et al.(2010b)]{pontoppidan2010b} Pontoppidan, K.~M., Salyk, C., Blake, G.~A., K\"{a}ufl, H.~U.\ 2010, \apjl, 722, L173 

\bibitem[Rothman et al.(1998)]{rothman1998} Rothman, L.~S., et al.\ 1998, \jqsrt, 60, 665

\bibitem[Salyk et al.(2011)]{salyk2011} Salyk, C., Pontoppidan, 
K.~M., Blake, G.~A., Najita, J.~R., \& Carr, J.~S.\ 2011, arXiv:1104.0948 Salyk, C., et al. \apj, submitted 2011

\bibitem[Salyk et al.(2008)]{salyk2008} Salyk, C., Pontoppidan, K.~M., Blake, G.~A., Lahuis, F., van Dishoeck, E.~F., \& Evans, N.~J., II 2008, \apjl, 676, L49

\bibitem[Shevchenko et al.(1991)]{shevchenko1991} Shevchenko, V.~S., Yakulov, S.~D., Hambarian, V.~V.,  \& Garibjanian, A.~T.\ 1991, \azh, 68, 275 

\bibitem[Sturm et al.(2010)]{sturm2010} Sturm, B., et al.\ 2010, \aap, 518, L129 

\bibitem[Thi \& Bik(2005)]{thi2005} Thi, W.-F., \& Bik, A.\ 2005, \aap, 438, 557

\bibitem[van Dishoeck \& Dalgarno(1984)]{vandishoeck1984} van Dishoeck, E.~F., \& Dalgarno, A.\ 1984, \icarus, 59, 305 

\bibitem[Willacy \& Woods(2009)]{willacy2009} Willacy, K., \& Woods, P.~M.\ 2009, \apj, 703, 479 

\bibitem[Yoshino et al.(1996)]{yoshino1996} Yoshino, K., Esmond, J.~R., Sun, Y., Parkinson, W.~H., Ito, K.,\& Matsui, T.\ 1996, \jqsrt, 55, 53 

\end{thebibliography}
\end{document}